\documentclass[%
      aps,
      prb, 
      reprint,
      twocolumn,
      superscriptaddress,
      showpacs]{revtex4-1}
      
\usepackage[utf8]{inputenc}
\usepackage[T1]{fontenc}
\usepackage{dcolumn}
\usepackage{bm}					
\usepackage{graphicx}
\usepackage{amsmath,amsfonts,amssymb,mathtools}	
\usepackage{latexsym} 				
\usepackage{xcolor}
\usepackage[caption = false]{subfig}
\usepackage[breaklinks, linkcolor = black]{hyperref}
\usepackage{tikz}
\usepackage{tabularx}
\usepackage{braket}
\usepackage{diagbox}
\usepackage{ulem} 
\usepackage{textcomp}
\usepackage{soul}
\usepackage[mathlines]{lineno} 
\usepackage{mathptmx}
\usepackage{blindtext}
\usepackage{booktabs,eqparbox}
\usepackage{siunitx}

\begin{document}

\title{Reorganization energy and polaronic effects of pentacene on NaCl films}

%

\author{Daniel \surname{Hernang\'{o}mez-P\'{e}rez}}

\email[e-mail:]{daniel.hernangomez@weizmann.ac.il}

\affiliation{Institute of Theoretical Physics, University of Regensburg, 93040 Regensburg, Germany}

\affiliation{Department of Materials and Interfaces, Weizmann Institute of Science, Rehovot 7610001, Israel}

\author{Jakob Schl\"or}

\affiliation{Institute of Theoretical Physics, University of Regensburg, 93040 Regensburg, Germany}

\author{David A. Egger}

\affiliation{Institute of Theoretical Physics, University of Regensburg, 93040 Regensburg, Germany}

\affiliation{Department of Physics, Technical University of Munich, 85748 Garching, Germany}

\author{Laerte L. Patera}
\affiliation{Institute of Experimental and Applied Physics, University of Regensburg, 93040 Regensburg, Germany}

\author{Jascha Repp}
\affiliation{Institute of Experimental and Applied Physics, University of Regensburg, 93040 Regensburg, Germany}

\author{Ferdinand Evers}

\email[e-mail:]{ferdinand.evers@ur.de}

\affiliation{Institute of Theoretical Physics, University of Regensburg, 93040 Regensburg, Germany}




\begin{abstract}
Due to recent advances in scanning-probe technology, 
the electronic structure of individual molecules can now also be investigated
if they are immobilized by adsorption on non-conductive substrates. 
As a consequence, different molecular charge-states are now  experimentally accessible.
Thus motivated, we investigate as an experimentally relevant example the 
electronic and structural properties of a NaCl(001) surface with and without 
pentacene adsorbed (neutral and charged) by employing density functional theory. 
We estimate the polaronic reorganization energy to be  
$E_\textnormal{reorg} \simeq 0.8-1.0$ eV, 
consistent with experimental results obtained for molecules of similar size.
To account for environmental effects on this estimate, different models for 
charge screening are compared. 
Finally, we calculate the density profile of one of the frontier orbitals 
for different occupation and confirm the
experimentally observed localization of the charge density upon \textcolor{black}{charging and} relaxation of \textcolor{black}{molecule-insulator interface} from \textit{ab-initio} calculations. 
\end{abstract} 
 
\maketitle

\section{Introduction}\label{sec:intro}
 
\begin{figure}
        \includegraphics[width=0.75\linewidth]{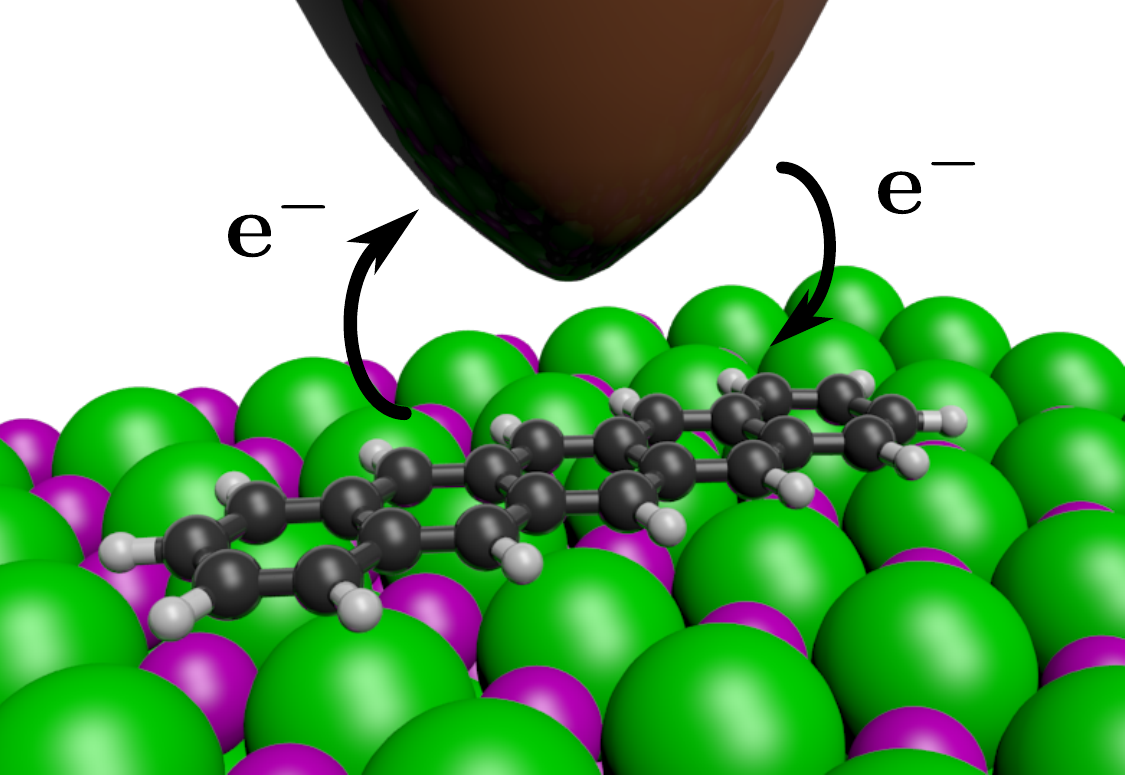}      
    \caption{Schematic representation of a scanning-probe 
    setup for single-electron transfer between a metallic tip
    and a pentacene molecule adsorbed on a NaCl film
    (violet: Na$^+$, green: Cl$^-$). 
    Single-electron transfer between the tip and 
    the molecule 
    are indicated with black arrows.
    }\label{f1} 
\end{figure}

The interaction of individual molecules with surfaces and interfaces constitutes one of the central topics of surface science since this field was launched. In the past, most of the interest focused on metallic surfaces, partially because of the relevance to catalysis, but also because experimental techniques such as the scanning tunneling microscope (STM) could not operate on non-conductive substrates. 
It was only relatively recently that this situation changed and STM-related techniques also became available  for individual molecules adsorbed on insulating films, {\color{black} which in turn are deposited on a metallic substrate}\cite{Repp2005}; for an illustration see Fig. \ref{f1}. Since then scanning-probe studies of individual molecules on insulating {\color{black} films}, such as sodium chloride (NaCl), have received a growing amount of attention (see, e.g., Refs.  [\onlinecite{Repp2006, Repp2013, Gross2014, Gross2015, Majzik2016, Hollerer2017, Hurdax2020}] and references therein). 
The main reason is that molecules tend to hybridize much more weakly with insulators than with metals and therefore new physical regimes become available. If sufficiently thick insulating substrates are used, the electron transfer from the adsorbed atoms or molecules is quenched and they can exhibit different metastable (long-lived) charge-states.\cite{Meyer2015} 
\textcolor{black}{Combining of STM techniques} with atomic force microscopy, 
\textcolor{black}{has allowed} us to experimentally obtain the energy gain of the system after charging a single-molecule\cite{Fatayer2018}, analyze the differences in the orbital shape with angstrom resolution\cite{Patera2019} or measure geometrical 
changes of single-molecules induced by electron transfer\cite{Fatayer2019}
Partially motivated by these experimental developments, 
calculations of the physical properties of atomic and molecular adsorbates
on insulating substrates have been attempted\cite{Olsson2007, Martins2010, Scivetti2013, Scivetti2014, Scivetti2017}. 
These calculations are very challenging due to 
the large number of electrons (especially from the 
{\color{black} 
substrate} layers)
involved.
Most notably, only approximate density functional theory (DFT) calculations\cite{Scivetti2013, Scivetti2014, Scivetti2017} 
have been performed %
\textcolor{black}{for single molecules adsorbed on insulating films. 
The additional simplification in these calculations is achieved by not explicitly including}
the metallic layer 
\textcolor{black}{which is replaced by a perfectly conducting ideal plane.}
The electrostatic interaction between the localized charge in the molecule and its image charge is self-consistently included
 by considering the energy of two charge distributions
separated by an homogeneous insulating film.
%

In this paper, we present an extensive computational study of a single
pentacene 
molecule adsorbed on a NaCl surface.
We focus on the understanding of the interplay between charging the 
molecule and the structural relaxation of the substrate. To address this
issue, we first comprehensively characterize the free surface (NaCl slab).
Already this arrangement, as it turns out, is nontrivial because, as we
show, the crystal structure of the bulk lattice is reached only very slowly with
increasing the slab thickness.

Depositing a single neutral pentacene molecule alters the alternating free surface 
by shifting away the nearby Cl$^-$ ions
toward the inside of the film by $\sim 8.5\%$ of the bulk lattice spacing. %
Upon adding an electron to the molecule, 
the main structural effect on the surface is that nearby Na$^+$ ions 
are attracted and bulge out towards the molecule
at the expense of the Cl$^-$ ions that go deeper inside.  
Importantly, the charge state of the molecule is experimentally found to be 
metastable if the NaCl-film with sufficient thickness is 
deposited on a copper substrate.
We estimate $0.8-1$ eV for the corresponding reorganization energy. 
To illustrate the effects of screening at the underlying metallic substrate 
different screening models have been compared. 

To connect to previous 
experimental work,
scanning tunneling profiles along the charged molecule have been simulated. 
We find that the charge-induced \textcolor{black}{molecule-insulator interface} relaxation 
enhances the tunneling probability around 
the center of the molecule. 
Our computational results are consistent with 
published experimental data.\cite{Patera2019}

\section{Model and methods}\label{sec:methods}

\subsection{\textcolor{black}{General computational details}}\label{sec:computing}
Our computational study focuses on NaCl films. We 
perform slab calculations with 
periodic supercells with (001) surface orientation employing  
DFT. 
Our DFT calculations are performed with the FHI-\textsc{aims} 
package\cite{Blum2009} that implements the Kohn-Sham formulation 
of DFT using a localized basis set. 
We use a standard generalized-gradient approximation 
(the Perdew Burke-Ernzerhof [PBE] functional \cite{Perdew1996}) for the exchange-correlation
functional and incorporate scalar relativistic corrections
to the kinetic energy at the level of the 
zeroth-order regular approximation.
As a basis set, we have routinely employed the 
optimized ``light'' settings (approximately equivalent to
``double zeta'' quality).
We have checked the robustness of our most prominent observations   
against ``tight'' settings (``double zeta + polarization'' quality).

Our standard convergence criteria are
$10^{-5}$ \textcolor{black}{electrons}/\AA$^{3}$ %
for the density, $10^{-3}$ eV for the sum of the Kohn-Sham eigenvalues and $10^{-6}$ eV for the total energy. 
The geometry optimization of the atomic structures start from an initial state in
which the atoms and ions are located relatively to their bulk (for the NaCl slabs) or slab 
(for pentacene on NaCl) positions.
They are performed using the energy-based Broyden-Fletcher-Shanno-Goldfarb optimization 
algorithm\cite{Blum2009,Press2007} and atomic structures are relaxed until 
every component of the residual force per atom or ion drops below the threshold $10^{-2}$ eV/\AA.
\textcolor{black}{For organic-inorganic interfaces, the account of dispersive
forces in DFT is known to be very important\cite{Scheffler2010, Maurer2016, Maurer2019}.}
We treat dispersive forces [van der Waals (vdW) interactions] using the Tkatchenko-Scheffler (TS) 
approach\cite{Tkatchenko2009} and the Clausius-Mosotti correction for solids\cite{Zhang2011} (TS-CM). Unless stated otherwise, we ignore dispersive forces inside the NaCl slabs, {i.e.} between Na$^+$-Na$^+$, Cl$^-$-Cl$^-$ and Na$^+$-Cl$^-$, but allow vdW interactions between all the other atomic species.
%
\begin{figure}
        \includegraphics[width=1.0\linewidth]{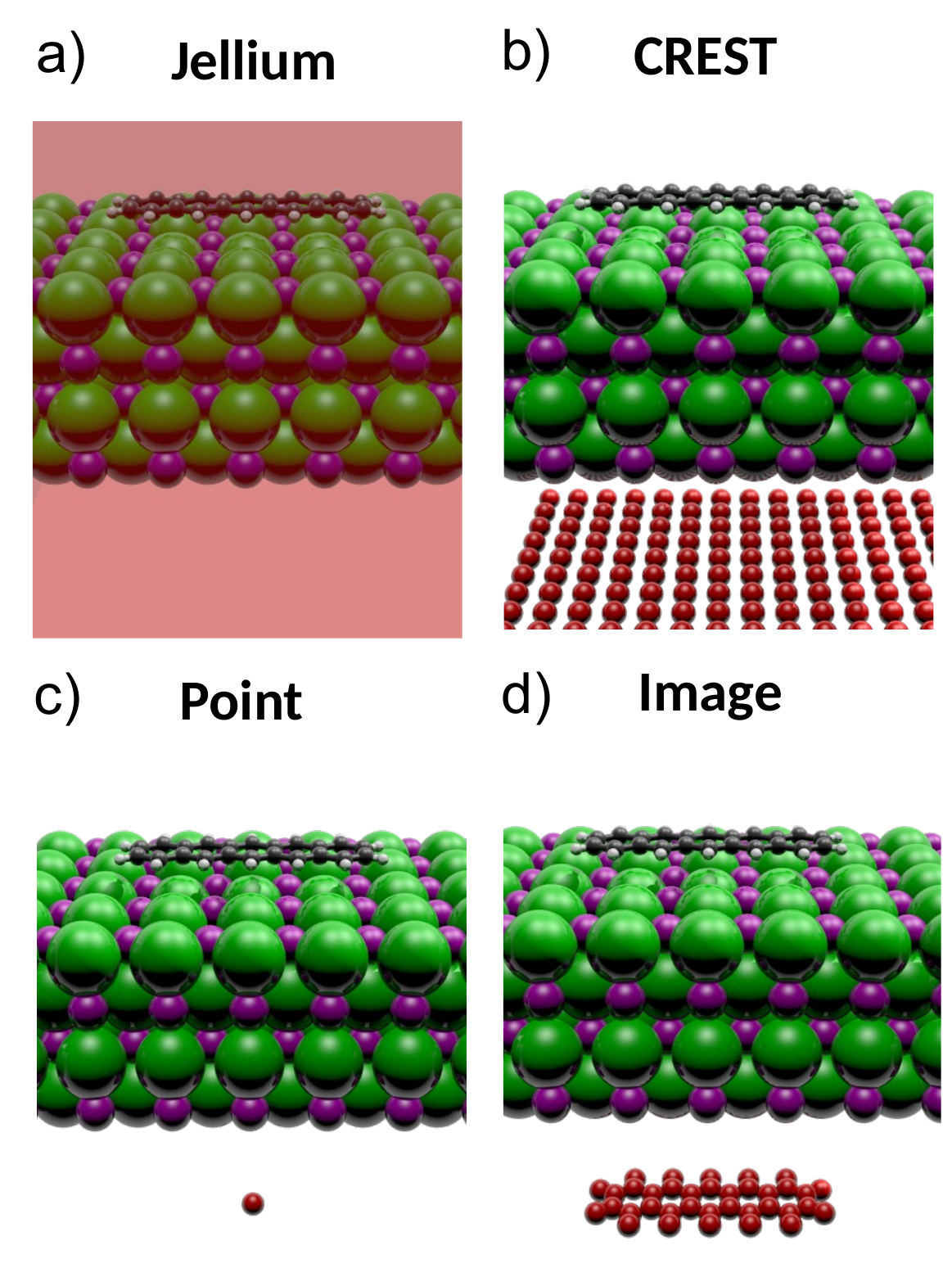}      
    \caption{Schematic illustration of the four different charge compensation methods  for periodic surfaces with excess
    charge used in this work. (a) Jellium model, here the transparent box represents the homogeneously charged background. From (b)-(d)
    the supercell cell counter charges are depicted by red spheres arranged as used (b) for CREST, (c) for the point charge, and 
    (d) for the shaped image charge.
    } \label{fS1}
\end{figure}
%

\subsection{Models for metal substrate}\label{sec:models_for_metal_surface}
As mentioned in Sec. \ref{sec:intro},
including in addition to the NaCl film a metallic layer in \textit{ab-initio} calculations is computationally prohibitive. 
We note that the main physical effects of these metal layers is to ensure, via screening, charge neutrality when charged atoms or molecules are adsorbed on the substrate. This behavior can be conveniently included in simplified models. We compared here some of those models for the NaCl slab with a charged pentacene adsorbed.

\subparagraph*{Jellium-model.} 
The simplest approach employs the jellium model, which ensures charge neutrality by adding a uniformly charged compensating background to the supercell\cite{Hofmann2019}, see Fig. \ref{fS1} (a). 
In band-structure codes,
the model is often implemented by omitting 
the zero-wavenumber component of the Coulomb interaction\cite{Deserno1998, Delley1996}. 
For slab calculations, the background extends also into the vacuum layer. 

As a consequence, the unit cell carries an electric dipole,
while the shape of the associated field is a modeling artifact
In the literature, two methods have been proposed to compensate for this: (i) extrapolation to the dilute limit by finite-size scaling\cite{Makov1995} and 
 (ii) \textit{a posteriori} corrections based on dielectric models\cite{Freysoldt2009, Komsa2012, Cao2017}. %

 \subparagraph*{Shaped counter-charge distribution.} 
 The artifact of the jellium model results from distributing the compensating background charge (``counter charge'') homogeneously over the unit cell. 
 More realistic arrangements are easily possible, though not standard in every band structure code. 
 Alternative choices are to 
 arrange the counter-charges as a two-dimensional sheet, Fig. \ref{fS1}(b), to contract them into a point, Fig. \ref{fS1}(c) or any intermediate geometry, Fig. \ref{fS1}(d), which can mimic the form of the image charge. 
On the technical level, the counter charges can be modeled as an ensemble of $N_\textnormal{counter}$ ion cores, with an individual fractional charge $q_\text{counter}$ each, and with a total charge matching the charge of the surface layer $Q$,  {i.e.} $Q = q_\textnormal{counter} N_\textnormal{counter}$.

We mention that the sheet geometry, Fig. \ref{fS1} (b), was adopted by Sinai \textit{et al.}
\onlinecite{Sinai2015} to describe surfaces of doped semiconductors, where the transfer of free charge carriers between the doped material and the surfaces leads to a space-charge region. In modeling, this region was contracted to a thin sheet that produces the associated electric field - hence the name charge-reservoir electrostatic sheet technique (CREST).

\section{Dimerization effects on pristine NaCl films} 
The reference geometry for our investigation on the effect of adsorbates on insulating substrates will be a clean NaCl(001) film. 
We therefore first extensively characterize the atomic geometry of this system. 

\begin{figure}
        \includegraphics[width=0.45\linewidth]{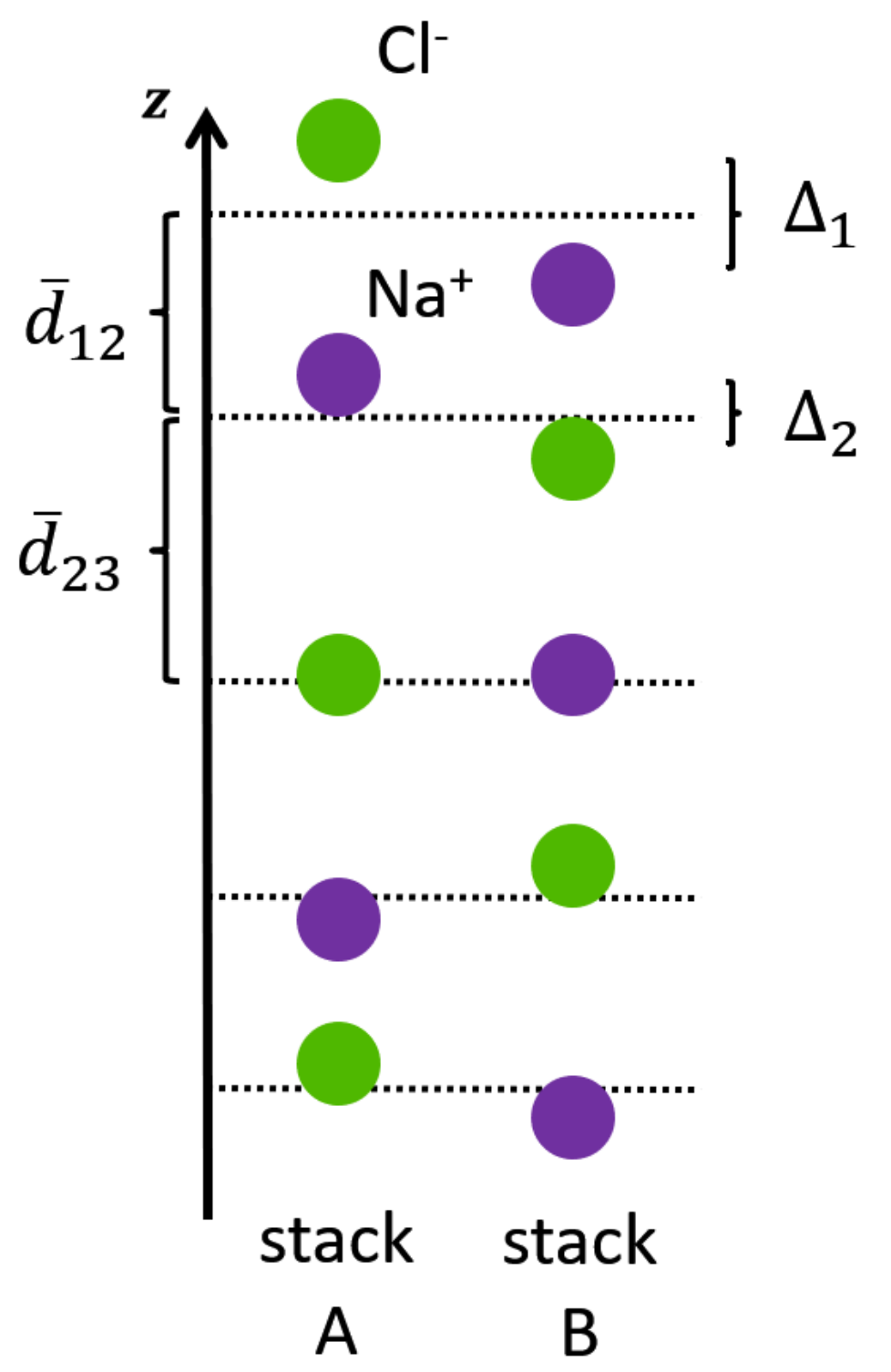}      
    \caption{Characterization of the surface relaxation in a NaCl(001) film.
    The mean layer separation and intralayer buckling of the first two
    layers of a NaCl supercell are shown.
    }\label{f2} 
\end{figure}
\subsection{Atomic geometry and relaxation}
The NaCl slab consists of two types of ionic slabs, denoted $A$ and $B$, showing
an alternation of Na$^+$ and Cl$^-$ ions along the $\hat{\mathbf{z}}$ direction 
(see Fig. \ref{f2}).
The ``surface relaxation'' describes 
the structural change of the crystal lattice 
that occurs due to the existence of a vacuum interface.
This change is characterized by 
the displacement of neighboring 
ions from each other measured relative 
to the bulk position; \cite{Vogt2001, Li2007} 
specifically we consider the {\it mean layer separation} 
\begin{equation}
 \bar{d}_{ij} = \dfrac{1}{2} \left[(z_i^A - z_j^A) + (z_i^B - z_j^B)  \right],
\end{equation}
where $z_i^{A/B}$ is the $z$-coordinate of ion $i$ in the 
stack $A/B$, and the {\it intralayer buckling}
\begin{equation}
 \Delta_i := z_i^A - z_i^B.
\end{equation}
In our calculations, only the relaxation normal to the surface has been accounted for, while the transverse coordinates are bulk-like and not relaxed. 
Lateral relaxations are not expected from symmetry considerations.

\subparagraph*{Computational details and models.}
Our initial geometries for the NaCl(001) slabs are constructed from converged bulk geometries (see Appendix \ref{app:bulk}). 
For NaCl(001) slabs, our DFT calculations and geometry optimizations
follow the convergence criteria for the SCF cycle detailed above with slightly
tighter force convergence parameter ($0.005$ eV/\AA).
We consider a vacuum spacing layer of $20$\,\AA \,(for which total 
energies are converged with a precision of $0.8$ meV
{\color{black} per layer}). 
We have verified that the 
structures are converged with a precision of $0.1$ pm
and also performed
numerous checks of the convergence with respect to the vacuum spacing and $k$-grid discretization (see Appendix \ref{app:checks}).

\begin{figure}
        \includegraphics[width=1.0\linewidth]{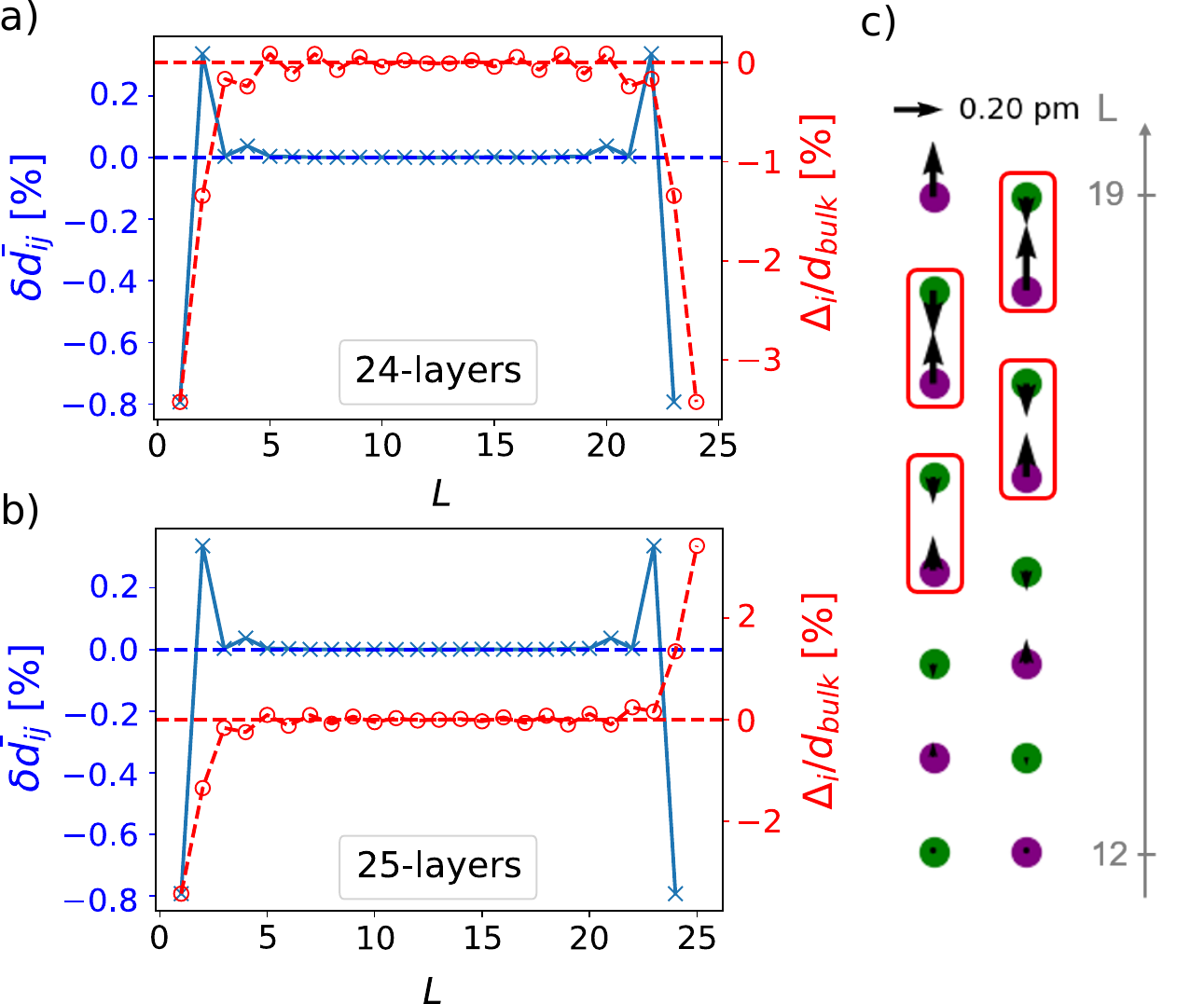}      
    \caption{(a) Relative mean layer separation (left, blue crosses) and intralayer buckling (right, red circles) measured with respect to the bulk interlayer separation for a freestanding relaxed NaCl slab of $24$ layers. Horizontal solid lines correspond to the reference value of the mean layer separation and buckling
    for the bulk crystal. 
    (b) Same as in (a) but for a $25$ layered slab. (c) Displacement of the ions
    from their bulk position upon structural relaxation. Dimerization of NaCl ions in a given stack 
    is indicated by the red boxes.
    }\label{f4} 
\end{figure}

\begin{table*}
	\begin{center}
	\setlength{\tabcolsep}{10pt} 
	\begin{tabular}{l|cc|ccc}
	\toprule
							& $\delta \bar{d}_{12}$ [\%]	& $\delta \bar{d}_{23}$ [\%] 	& $\Delta_1/d_\textnormal{bulk}$ [\%] & $\Delta_2/d_\textnormal{bulk}$[\%]&  $\Delta_3/d_\textnormal{bulk}$[\%]  \\
		\hline
		PBE	(12 lay.)			& -0.786				& 0.334					& -3.472 			& -1.297 			& -0.210 				\\
		PBE	(24 lay.)			& -0.793				& 0.337					& -3.437 			& -1.333 			& -0.175 				\\
		PBE-TS-CM (12 lay.)			& -0.255				& 0.362					& -3.924			& -1.490			& -0.327				\\
		PBE [\onlinecite{Li2007}]	& -0.600				& 0.020					& -3.860			& -1.053			& 0.000 				\\ 
		Experiment [\onlinecite{Vogt2001}] & -1.430				& 0.100					& $-5.0\pm 0.1$	& $-0.7\pm 0.1$	& $0.0 \pm 0.1$ \\ 
		\bottomrule
	\end{tabular}
	\caption{Mean layer separation in percentage of the bulk lattice spacing, $\delta \bar{d_{ij}} = (d_{ij} - d_\textnormal{bulk})/d_\textnormal{bulk}$, and intralayer buckling, $\Delta_i/d_\textnormal{bulk}$, for the three topmost layers of a free-standing relaxed NaCl(001) slab. 
For the NaCl(001) lattice, $d_\textnormal{bulk} = a_0/2$, where $a_0$ is
the bulk lattice constant (see bulk structural details in Appendix \ref{app:bulk}).
	The subscripts $1,2,3$ designate the top surface, second layer and third layer, respectively. We compare our results for the PBE functional ($L=12,24$) and PBE with TS-CM corrections ($L=12$) to previous \textit{ab-initio} calculations and experimental results obtained using low-energy electron diffraction $I-V$ analysis. The experimental value $d^{\textnormal{exp}}_\textnormal{bulk} = 2.8025$\AA\, is 
	estimated from the average value of the experimental lattice constants, see Table S1.
	}\label{t1}
	\end{center}
\end{table*}


%

Depending on the physical situation at hand, 
thin films - freely suspended or on substrates - or thick films, different boundary conditions are appropriate for modeling. 
For example, with freely suspended films the upper and the lower sides of the film can relax freely
(``symmetric slabs''); in contrast, when modeling the surface of a thicker film, the ions of one side of the slab can be fixed to the bulk values (``asymmetric slabs''). We have ensured in detailed separate calculations that our conclusions do not change for this asymmetric situation (not shown).

\subparagraph*{Mean layer separation.} 

Our main results are summarized in Fig. \ref{f4}. We compare two slabs with even ($L=24$) and odd ($L=25$) number of layers
to investigate even-odd effects. For comparison, in Table \ref{t1} we also reproduce results for $L=12$.
The data allows 
for the following conclusions: 

Qualitatively,
we observe that
the inter-layer distances oscillate when going from the surface into the bulk, see blue traces in Fig. \ref{f4}, (a) and (b).
We attribute the effect to the reduced coordination number of the surface atoms, which tends to strengthen the remaining bonds to the second layer.  The second layer in turn coordinates more strongly with the first layer and so binding to the third is less pronounced. 
This mechanism -- well-known from closed-packed metal surfaces -- results in an oscillating inter-layer spacing, 
with the amplitude of these oscillations decaying within few layers into the bulk.
To illustrate this, we compare in Table \ref{t1} the 
mean layer separation  of first-second and second-third layers for 
two different thicknesses, $L{=}12,24$. 
Quantitatively, the impact of surface relaxation on the mean layer separation is of the order of 0.5\% in terms of the bulk interlayer distance, $d_\textnormal{bulk}$, for the first and second layers and a factor of $2$ smaller for the second and third. 

\subparagraph*{Intralayer Buckling.}
Qualitatively, Cl$^-$ ions of the first layer relax outwards 
of the surface compared to the Na$^+$ ions.
There, the buckling is stronger, 
up to almost 4\% in units of $d_\textnormal{bulk}$ 
and has a negative sign, indicating that the Cl$^-$ ions always relax outwards while the Na$^+$ ions inwards
Together with the alternating stacking of ions in NaCl 
this behavior is a manifestation of \textit{dimerization} in NaCl where Na$^+$ and Cl$^-$ ions come closer in alternating way along $\hat{\mathbf{z}}$. We show a sketch of the dimerization occurring in the central
region of the slab in Fig. \ref{f4} (c). The NaCl dimers originated by surface creation are marked with red rectangles.
As compared to the mean layer separation, the buckling and the 
associated dimerization decay much slower:
$\Delta/d_\textnormal{bulk}$ keeps a value of $\sim 0.1$\% 
even $10$ layers deep into the bulk of the slab.
Indeed, we have checked (not shown here)
that slabs of up to $50$ layers still present 
dimerization between Na$^+$ and Cl$^-$ ions along a stack.

\subparagraph*{Relation to experiments and earlier DFT studies.} 
Finally, as inferred from Table \ref{t1} 
our results are in good qualitative agreement with experimental measurements, with
quantitative deviations that can reach a factor of 2-3. 
Comparing to earlier computational work\cite{Li2007}, 
we report a relatively large discrepancy 
of around $24\%$ difference (roughly $0.3~\si{pm}$) for $\delta \bar{d}_{12}$ and 
$94\%$ difference (roughly $0.6~\si{pm}$) for $\delta \bar{d}_{23}$.
We attribute these deviations to methodological differences in DFT-implementations 
of FHI-aims (all electron, localized basis set) versus 
CASTEP (pseudopotentials, plane waves). 

\subparagraph*{Discussion of functional dependencies.} One infers from the data given in Table \ref{t1} that the \textit{ab-initio} obtained geometries have a residual dependency on the DFT functional. 
Specifically, we have tested the effect of dispersive interactions; our numerical
data in Table \ref{t1} shows they can have a significant quantitative but no qualitative effect. 
\footnote{For PBE, the interlayer distance $|\delta \bar{d}_{12}|$ is larger
than $|\delta \bar{d}_{23}|$. 
Surprisingly, this is not the case
when including the TS-CM dispersion.} 

\subsection{Surface energy} 

We turn to the investigation of the surface energy per area, $\gamma$. 
It derives from the scaling of the total energy of a slab with 
$L$ layers: 
\begin{equation}
 \dfrac{E_\textnormal{slab}(L)}{L} -
 E_\textnormal{bulk} \coloneqq \gamma A \dfrac{1}{L} + \Delta.
 \label{e3} 
\end{equation}
Here, $A$ denotes the surface area of the slab unit cell
and $E_\textnormal{bulk}$ is obtained from a separate
bulk calculation (see Appendix \ref{app:bulk} for technical details). 
The value for $\gamma$ can be obtained fitting the numerical data as
a function of $1/L$. The parameter $\Delta$ is kept as a 
sanity check; correct fitting must result in an estimate 
for $\Delta$ consistent with zero. 
%

\begin{figure}
        \includegraphics[width=1.0\linewidth]{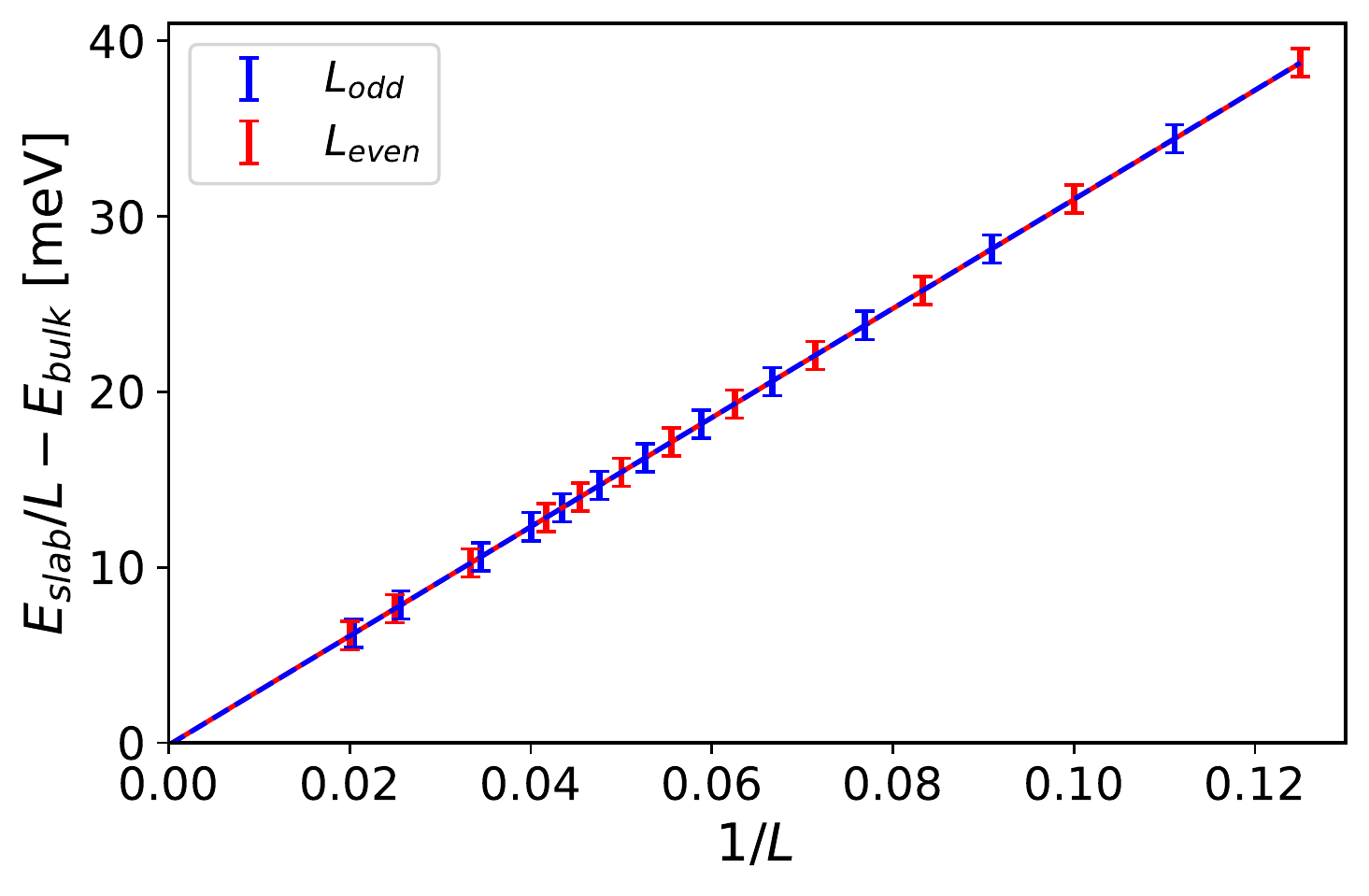}      
    \caption{Slab energy per layer, $E_\textnormal{slab}(L)/L$,  as a function of the inverse
    number of layers, $1/L$ for odd-numbered (blue) and even-numbered (red) free standing NaCl films. The bulk energy offset is equal
    to \textcolor{black}{$E_\text{bulk}=-17006033.751~\si{meV}$}.
    }\label{f6} 
\end{figure}

We show the numerical results for odd- (red line) and even-layered (blue line) slabs in Fig. \ref{f6}. 
Fitting to Eq. \eqref{e3} yields parameters reproduced in Table \ref{t2}. 
As one would expect, in the limit of thick slabs both yield 
identical results within the error bars (given by the accuracy of the total energy with respect to the vacuum spacing). 
Our results for $\gamma$ are in excellent agreement with 
previous DFT calculations\cite{Li2007}, which give a substantially broader
interval $9-10$ meV/\AA$^2$.

\begin{table}
	\centering
		\setlength{\tabcolsep}{10pt} 

\begin{tabular}{l|cc}
	\toprule
		& $\gamma$ [$\si{meV}/{\text{\AA}^2}$] 	& $\Delta$ [$\si{meV}$] 	\\
		\midrule
		odd	& 9.6 $\pm$0.2		& -0.1 $\pm$0.4		\\
		even	& 9.6 $\pm$0.2		& -0.1 $\pm$0.4	\\
		\bottomrule
			\end{tabular}
	\caption{Surface energy, $\gamma$, and offset parameter, $\Delta$, obtained for even and odd number of layered slabs. As the surface area of the slab supercell we used the DFT-optimized value, $A  = 16.256$ \AA\textsuperscript{2}.}
	\label{t2}
\end{table}


%

\section{Molecule-induced surface reorganization}\label{sec:molecule}

We turn to the investigation of 
the NaCl surface reorganization after physisorption  of 
a single pentacene molecule.

\subsection{Adsorption energy and binding geometry}

\begin{figure}
        \includegraphics[width=1.0\linewidth]{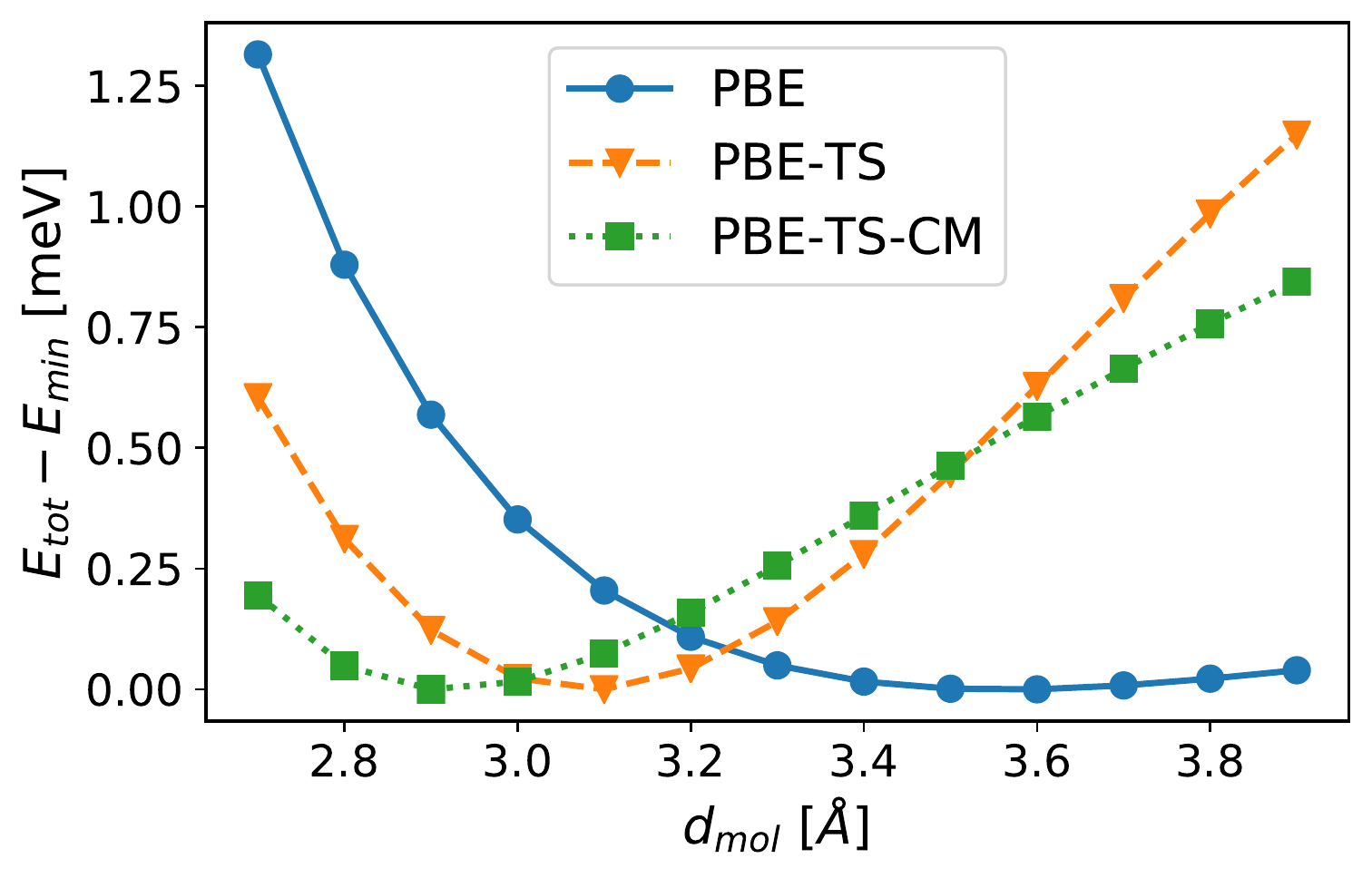}      
    \caption{Binding \textcolor{black}{potential} of pentacene \textcolor{black}{adsorbed} on a NaCl(001) surface \textcolor{black}{as a function of} the distance \textcolor{black}{between} the molecule \textcolor{black}{and} the surface. DFT-calculations employ PBE without (blue circles) and with vdW-correction (PBE-TS, orange triangles; PBE-TS-CM, green squares). Energies,  $E_\textnormal{tot}$, are given relative to the 
    energy of the fully relaxed system, $E_\textnormal{min}$.
    }\label{f9} 
\end{figure}

\textcolor{black}{We first report on the adsorption properties of neutral pentacene on NaCl}. From a DFT-relaxed structure we extract the optimal adsorption distance, $d_\textnormal{opt}$, and the adsorption 
energy, $E_\textnormal{ads}$, computed 
from the total energy as
\begin{equation}
E_\text{ads} = E_\text{P-NaCl}(d_\textnormal{opt}) - E_\text{NaCl} - E_\text{P}.
\end{equation}
Here, $E_\text{Pc-NaCl}$ is the energy of pentacene adsorbed on NaCl, $E_\text{NaCl}$ 
the energy of the NaCl slab and $E_\text{P}$ the energy of isolated pentacene.
The binding potential, \textcolor{black}{Fig. \ref{f9}}, is obtained from 
the total energy by varying the distance between 
the molecule and the surface of the NaCl slab without relaxation of the ionic coordinates.
We monitor directly the atomic geometry (nuclear coordinates) of relaxed pentacene on NaCl after geometry optimization. 

\begin{table}
	\centering
		\setlength{\tabcolsep}{10pt} 
		
	\begin{tabular}[b]{l|cc}
	\toprule
		Functional				& $E_\text{ads}$ [\si{eV}]	& $d_\text{opt}$ [\AA]	\\
		\midrule
		PBE 						& -0.25 	& 3.56	\\
		PBE-TS 						& -2.71 	& 3.12 	\\
		PBE-TS-CM 					& -1.55 	& 2.95	\\
		optB86b-vdW [\onlinecite{Scivetti2017}]	& -1.65 	& 3.05	\\
		Experiment [\onlinecite{Repp2005}] 
		& - & \hspace{-.5cm}$\simeq 3.0$ \\
		\bottomrule
	\end{tabular}
	\caption{Adsorption energies, $E_\text{ads}$, and distances, $d_\text{opt}$, estimated experimentally and calculated with different functionals and vdW dispersion schemes. \textcolor{black}{For computational details, in particular, with respect to geometry optimization see main text.}
	}\label{t3}
	\end{table}


\subparagraph*{Additional computational details.} 
The geometry optimizations for pentacene on NaCl slabs are performed
in rectangular supercells 
of sizes $n \times m$ Cl$^-$ ions in the $x-y$ plane. 
We ensure that
the distance of the adsorbed
molecule to the boundaries of 
the supercell
is the same in both directions.
Geometry optimizations start from an initial geometry in which 
the gas-phase molecule and a four-layered NaCl 
slab are structurally relaxed
independently. The two bottom-most 
layers are kept frozen (``asymmetric slab''). 
Next, we place the molecule at a distance of $3.1$ \AA\,
from the top layer of the slab and perform the minimization
of the forces using DFT\cite{Blum2009} following the criteria detailed above. 
Dispersion forces are
described within the PBE-TS scheme.
For charged pentacene/NaCl systems, the counter-charge
distribution in the supercell is located at a fixed distance of 8 \AA\, from the bottom layer of the NaCl film.

\subparagraph*{Results for the binding properties.}
After geometry optimization, we find the central benzene ring of pentacene 
located on top of a Cl$^-$ ion and 
the molecular long symmetry axis oriented
along the polar direction of the 
surface (\textit{i.e.} along
a string of Cl$^-$ ions). 
Our computational results are consistent with experimental adsorption geometry\cite{Repp2005}.
The optimal geometry results from a competition of several forces including electrostatic, e.g. Hartree-type, the vdW  attraction and the Pauli repulsion. 
We rationalize our result in the following way: An important contribution from electrostatics stems from the 
interaction between the surface ions
and the charge distribution on the molecule.
%
Since most of the electronic charge lies on
the benzene rings,
the electrostatic repulsion can be minimized
by having a string of Cl$^-$ ions to be located along the long 
symmetry axis of the molecule and thus further apart from the C atoms.
This way, also a larger number of Na$^+$ ions are allowed to be closer to
the molecular charge on the C atoms. 
The collinear geometry is also favorable for reducing the Pauli repulsion. 
Here, the reason is that the ionic radii of Cl$^{-}$ ions are larger than Na$^{+}$
and therefore the Pauli repulsion occurs for Cl$^{-}$ at larger distances.
This then also favors Cl$^{-}$ to be located further away from the C atoms.

The quantitative estimates for the binding \textcolor{black}{potential} and optimal adsorption distance
are summarized in Table \ref{t3} and Fig. \ref{f9}.
Without vdW-corrections, we find 
an unrealistically flat binding 
curve with a small adsorption
energy, $-0.25$ eV, and a binding distance $d_\textnormal{opt}$ approximately $20\%$ bigger
than the experimental estimations.
Including vdW dispersion 
 increases the curvature of the binding  both for PBE-TS and PBE-TS-CM. The binding distance agrees
with experiment and previous DFT calculations using the optimized non-local exchange
correlation functional optB86b-vdW.\cite{Scivetti2017}
%
%

\begin{figure}
        \includegraphics[width=0.95\linewidth]{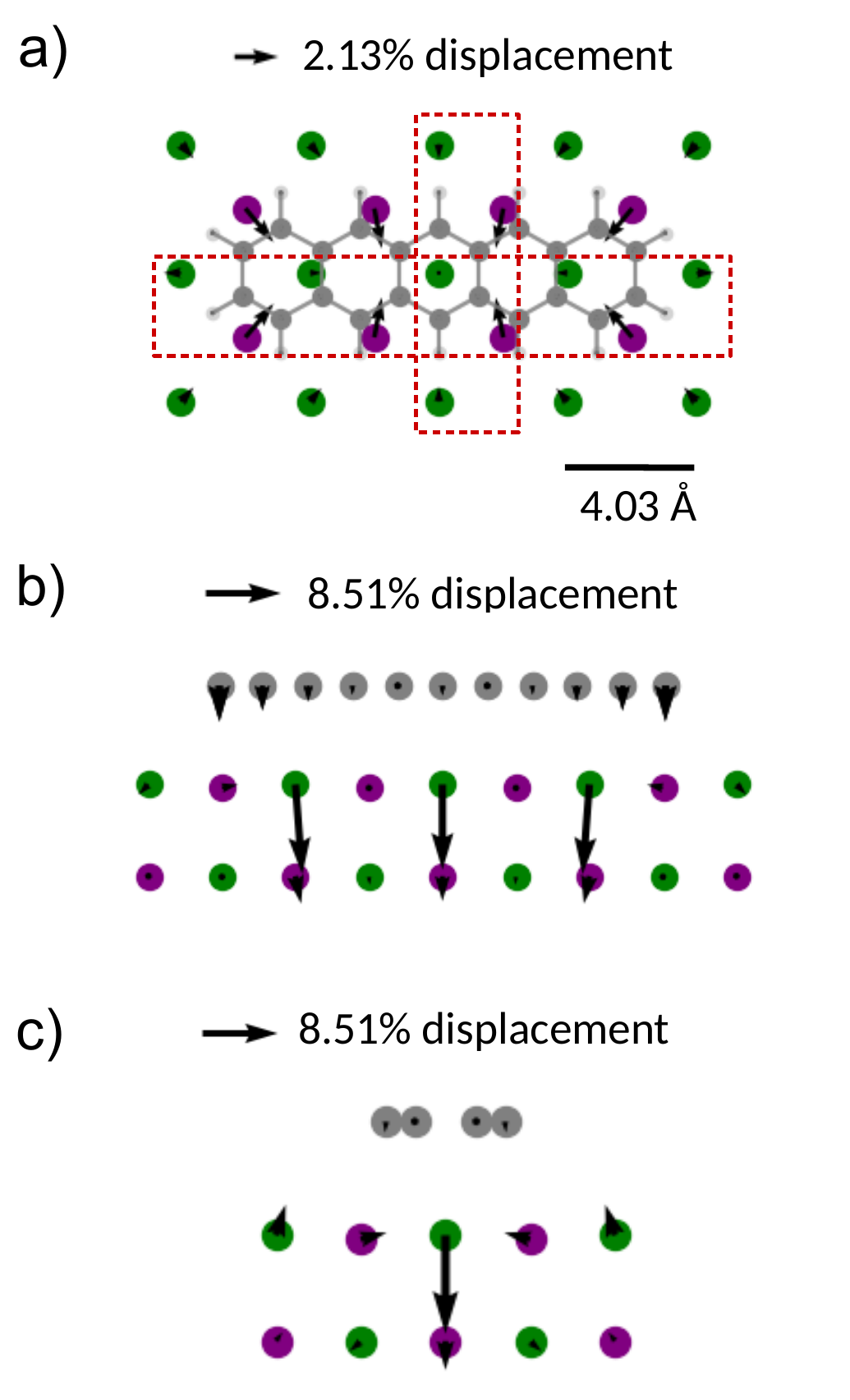}      
    \caption{Displacements of Na$^+$, Cl$^-$ ions and C atoms  
    upon adsorption of pentacene; only the first two top-layers out of four are displayed. Atomic displacements are shown for (a) the molecular and top surface layer in the $x-y$ plane, and red-boxed areas across (b) $x-z$ plane and (c) $y-z$ plane. The displacement is given as a percentage with respect to the \textcolor{black}{bulk interlayer distance}, $d_\textnormal{bulk}$, of the NaCl crystal. \textcolor{black}{The displacement arrows are scaled by a factor of 50 in panel (a) and a factor of 100 in panels (b) and (c).}
    }\label{f8} 
\end{figure}

\begin{figure}
        \includegraphics[width=0.905\linewidth]{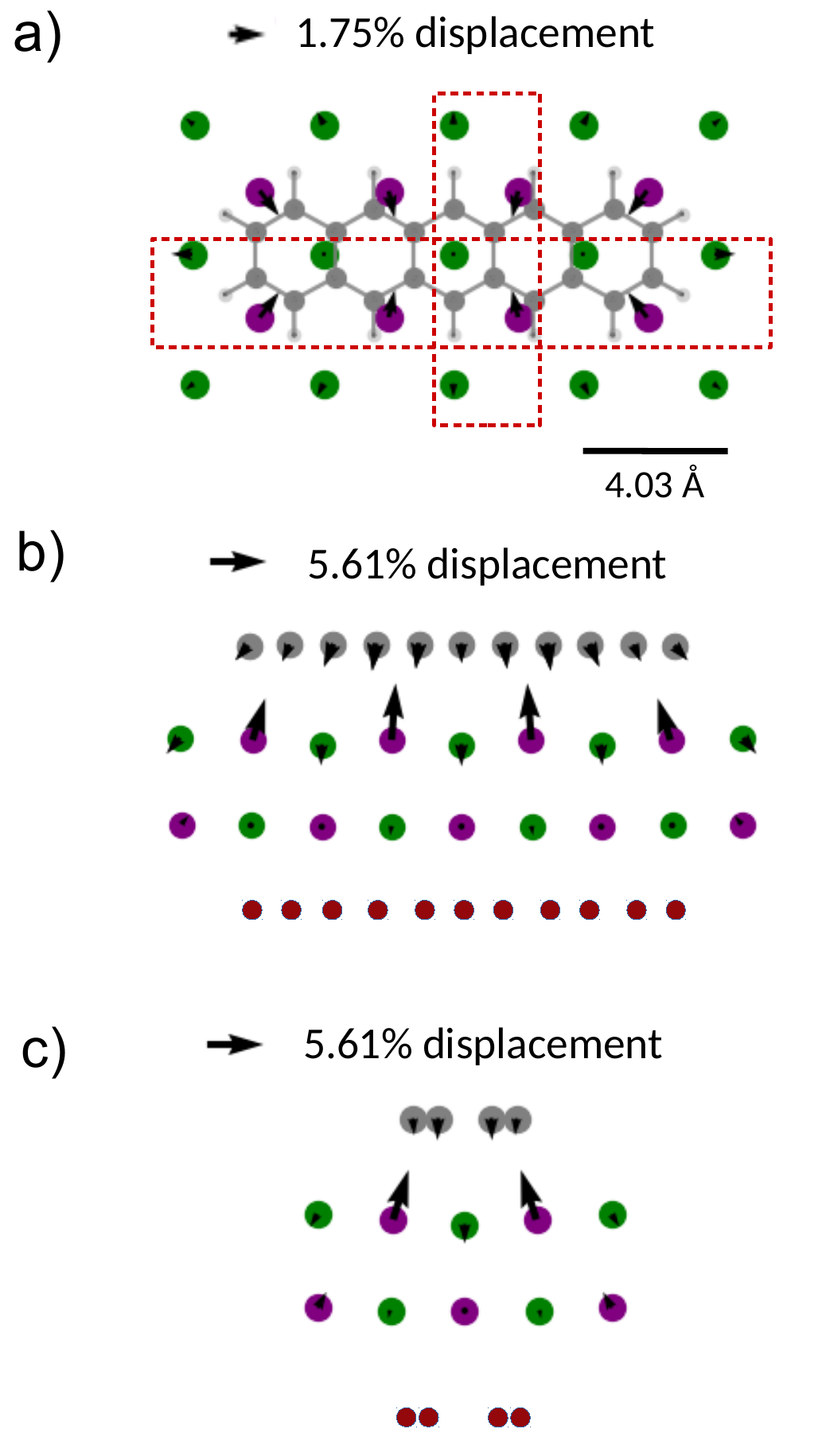}      
    \caption{Schematics analogous to Fig. \ref{f8} after charging pentacene with one electron; also the counter-charges (red discs) are indicated. 
   Atomic displacements are obtained from the difference between the charged geometry, $\textnormal{geo}^{-1}$
    to the neutral geometry, $\textnormal{geo}^0$ and given as a percentage of
   the interlayer distance, $d_\textnormal{bulk}$, of the NaCl bulk crystal. 
   The displacements
    are shown for (a) the molecular and top surface layer in the $x-y$ plane, and red-boxed areas across (b) $x-z$ plane and (c) $y-z$ plane. \textcolor{black}{The displacement arrows are scaled by a factor of 120 in panel (a) and a factor of 60 in panels (b) and (c).}
  }\label{f10} 
\end{figure}

\subsection{Detailed response of pentacene/NaCl \textcolor{black}{adsorption} geometry} 

While \textcolor{black}{adsorption} energies in Table \ref{t3}  are seen  to exhibit a rather strong functional dependency (by almost a factor of $2$), the binding geometries of the vdW-corrected calculations are rather close. In the following, we therefore use the computationally inexpensive PBE-TS treatment for pentacene/NaCl structure optimization and subsequently check our calculations with PBE-TS-CM. 


%
\textcolor{black}{We turn now to the investigation of the response of the 
NaCl film geometry upon adsorption of neutral and charged pentacene}.
In Fig. \ref{f8}, we show the atomic displacements
for a $7\times 5$ supercell cell size in the $x-y$ plane. As one would expect, most of the substrate reaction to the adsorption of the \textcolor{black}{neutral molecule}
occurs in the 
topmost layer of the NaCl film; 
the second layer shows a similar pattern with significantly smaller displacements.
The Na$^+$ ions are attracted by the molecule
and relax -- predominantly laterally -- by an amount of $\sim 6$~pm, i.e., 
roughly $2$\% of the NaCl layer spacing, $d_\textnormal{bulk}$.
The Cl$^-$ ions that are located along
the molecular long axis are 
pushed deeper into the substrate 
by $8.51\%$ (roughly $24.2\, \si{pm}$) 
[see Fig. \ref{f8} (b) and (c) respectively].
Note that this trend is opposite to the trend given by the buckling of the clean NaCl surface, as seen in Fig. \ref{f4}, thus partly ironing out the uppermost layer.
%
We also observe a small bending of 
outer rings in the molecule upon adsorption by $2$\% \textcolor{black}{of $d_\textnormal{bulk}$} (roughly \textcolor{black}{corresponding to $10^\circ$)}.

%
Upon charging the pentacene molecule dives deeper into the 
NaCl surface 
by 
$2\% \, d_\textnormal{bulk}$ ($\sim 6~\si{pm}$) due to electrostatic interaction.
\textcolor{black}{
The carbon-carbon bonds of outermost benzene rings parallel to the long symmetry axis of the pentacene molecule
increase their length by roughly $1$\%. Simultaneously, the carbon-carbon bonds of the central ring shorten by $\sim 0.4 \%$ ($\sim 0.5$ pm).
This results in an increase in the length of the molecule
of $\sim 8$ pm.
}
In addition,  \textcolor{black}{the trends observed in the substrate after} 
 adsorption of neutral pentacene are enhanced, see Fig. \ref{f10}: 
The excess charge attracts
the Na$^+$ ions even more, and now bulge out of the surface
towards the molecule.
The Cl$^-$ ions are pushed further downwards, inverting the sign of the buckling near the molecule.

\subparagraph*{{Methodological discussion.}}
For charged pentacene/NaCl systems, we have checked that
different models for the metal substrate
yield the same qualitative
response of the pentacene/NaCl surface after charging. 
The quantitative difference found is small: 
We find a maximum displacement of the 
ions
of $5.42\%$ for CREST, $5.73\%$ for point-shaped 
charge distribution and 
$4.92\%$ for the jellium-model
in units of $d_\textnormal{bulk}$.
For comparison, we get a maximum displacement of $5.61\%$ 
for the image-shaped charge distribution (shift of $\sim 16$ pm){\color{black}, see Fig. \ref{f10}}.
Maximal deviations in each layer parallel to the molecular
plane are small ($0.1$\%).
Ionic displacements between the metal models with
explicit charge compensation that create a realistic dipole
 inside the supercell agree better with each other 
 compared to the uniform background model.
Small quantitative differences between them can be readily understood
from  different
spatial distribution of the dipole fields.
%
%

\begin{figure}
        \includegraphics[width=1.0\linewidth]{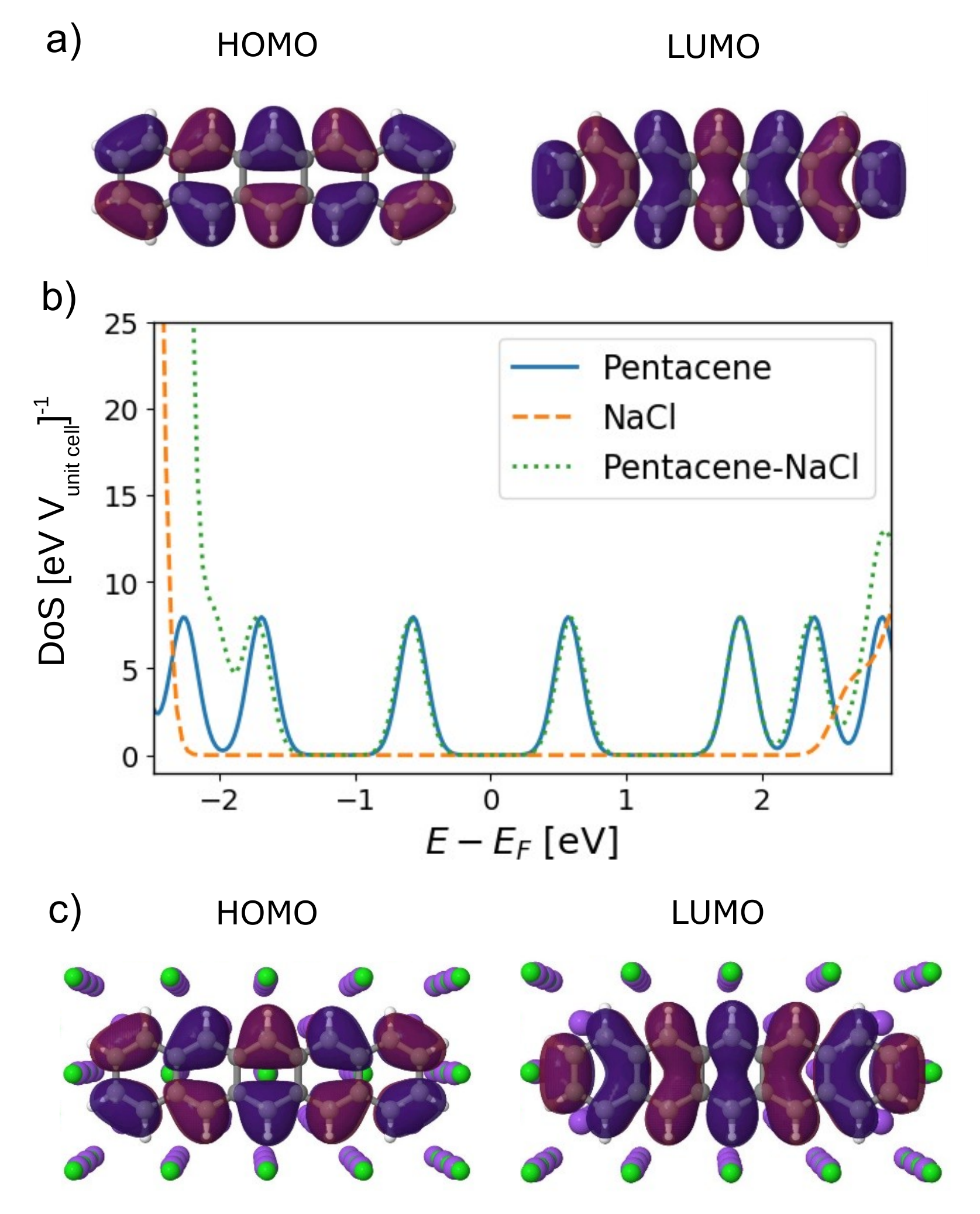}
    \caption{(a) \textcolor{black}{Highest occupied (HOMO, $-4.50$ eV) and lowest unoccupied (LUMO, $-3.35$ eV)} molecular orbital of the isolated pentacene molecule and 
    (b)
    corresponding density of states (DoS, blue solid curve). Also shown is the DoS of the 4-layer NaCl slab (orange dashed curve) and of pentacene adsorbed on NaCl (green dotted curve) in the NaCl band gap region.
    \textcolor{black}{For each DoS curve the energy is referred to the Fermi energy defined as $(E_\textnormal{LUMO} + E_{\textnormal{HOMO}})/2$ (and equal to $-3.92$ eV for neutral pentacene, $-4.51$ eV for NaCl and $-4.70$ eV for pentacene-NaCl).}
    The DoS peaks were broadened using Gaussians with broadening of \textcolor{black}{0.1} eV.
    (c) \textcolor{black}{HOMO ($-5.29$ eV) and LUMO ($-4.11$ eV)} of adsorbed pentacene molecule on NaCl.    %
    }\label{f7} 
\end{figure}

We also have investigated the effect of including CM corrections 
to the vdW dispersion at the PBE level (not shown). We 
report, both for neutral and charged pentacene/NaCl systems,
a residual dependency on the vdW-scheme.
The displacements out of
the molecular plane are larger by
 $\sim 1$\% (neutral) and $\sim 4$ \% (charged pentacene).
Parallel to the molecular plane, the atomic
displacements are very similar compared
to PBE-TS.
We also observe a stronger bending of the molecule
at both ends by
$4.87\%$ - roughly $13~\si{pm}$ for neutral adsorption.
Stronger finite size effects and surface reaction
are expected from the larger CM corrections to
the polarizability of the NaCl surface. 

\subsection{Electronic structure of pentacene adsorbed on NaCl}

%
%
The molecular density of states (DoS) and relevant molecular orbitals of neutral free pentacene and \textcolor{black}{neutral}
pentacene/NaCl are displayed in Fig. \ref{f7}.
The frontier orbitals highest occupied molecular orbital (HOMO) and lowest unoccupied molecular orbital (LUMO) 
in free and adsorbed pentacene exhibit 
the same nodal structure and similar spatial distribution 
 \textcolor{black}{
 \textcolor{black}{Furthermore}, the Kohn-Sham
 gap is almost insensitive to adsorption, $1.15$ eV for gas-phase pentacene versus  
 $1.18$ eV for pentacene/NaCl.
 \textcolor{black}{While these} values are in very good 
 agreement with previous DFT calculations\cite{Scivetti2017} 
 ($\pm 0.01$ eV),
 \textcolor{black}{note that they strongly disagree with}
 the experimental transport gap (which is
 measured close to $4.4$ eV \textcolor{black}{for 3-layered NaCl films}\cite{Repp2005}.) This is due to the missing derivative discontinuity\cite{Perdew1983, Sham1983} and absence of substrate-induced renormalization\cite{Neaton2006} in the exchange-correlation functionals employed here}.

Finally, we observe that the frontier orbitals lie within the large band gap of NaCl far from the conduction and band edges, 
as seen in the 
DoS, Fig. \ref{f7} (b). Therefore, as one would expect for an insulating substrate  \textcolor{black}{any} excess charge \textcolor{black}{is expected to remain} localized on the molecule.  

\subsection{Reorganization energy}

The reorganization energy
(also {\it polaronic shift}) quantifies the energy gain that results from the reaction of the pentacene/NaCl geometry upon charging. 
Experimentally, it can be accessed, {e.g.}, by means of the scanning probe setup  shown schematically in Fig. \ref{f1}. 
%
%
%
%
\begin{figure}
        \includegraphics[width=1.0\linewidth]{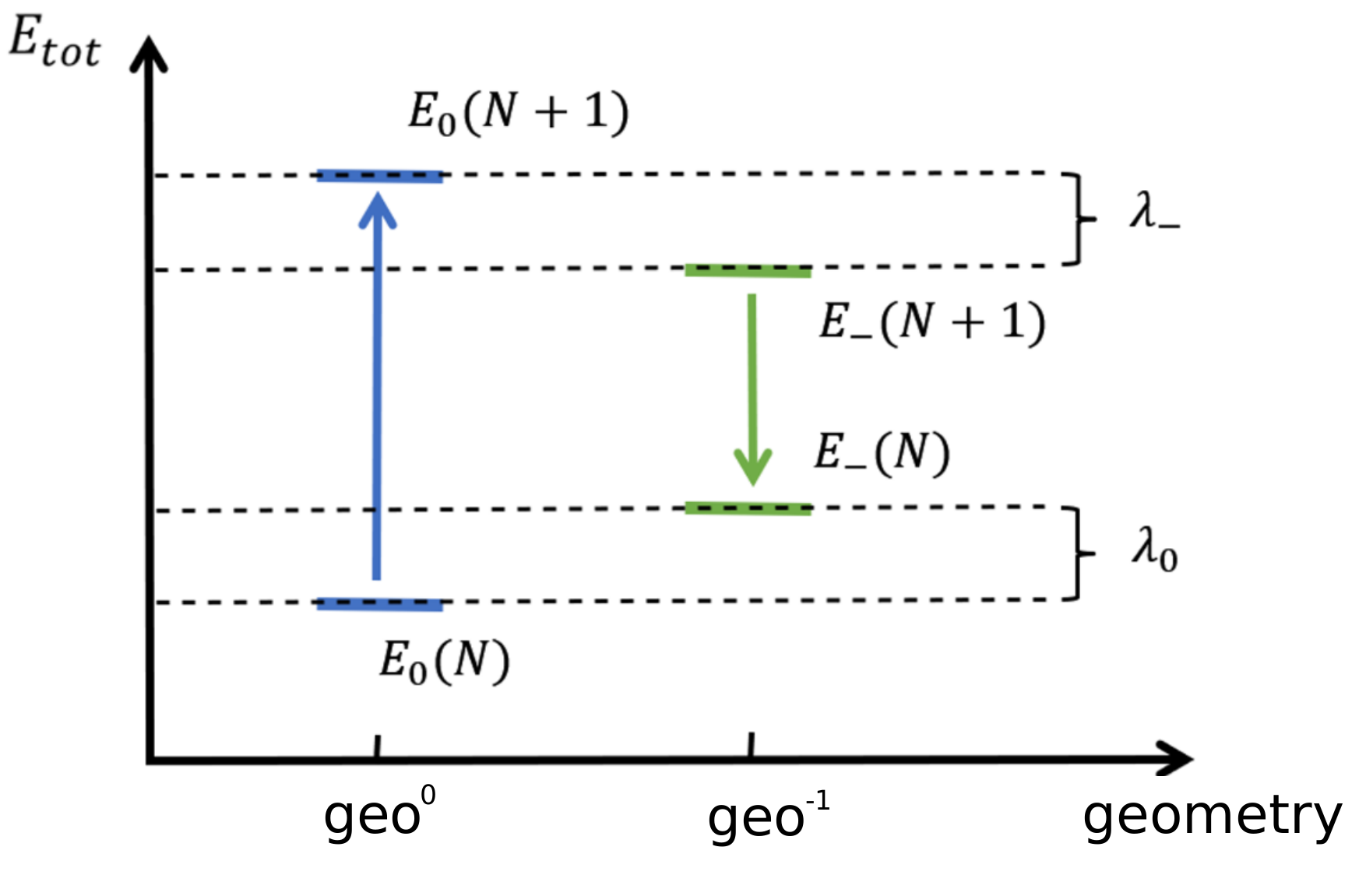}      
    \caption{Sketch representing the calculation of the reorganization energy
    using the total energy obtained from DFT. Here, $N$ denotes the total number of electrons, $\lambda_0$ and
    $\lambda_-$ represent the relaxation energies from the neutral (0) and charged ($-$) system.
    The $0 \rightarrow 1^-$ and $1^- \rightarrow 0$ many-body transitions which can be accessed experimentally\cite{Fatayer2018} 
    are shown by the blue and green vertical arrows, respectively.
    }\label{f3} 
\end{figure}
%
Formally, the reorganization energy can be defined as 
\begin{align}\label{eq:reorganization}
 E_\textnormal{reorg} := \left[E_{-}(N) - E_{0}(N)\right] - \left[E_{0}(N+1) - E_{-}(N+1)\right],
\end{align}
where $N$ denotes the number of electrons in the neutral system and the suffix indicate
 whether the total energy corresponds to the structurally relaxed charged
($-$) or uncharged ($0$) geometry.
The bracketed terms in Eq. \eqref{eq:reorganization} are identified as relaxation energies from the neutral, $\lambda_{0} := E_{-}(N) - E_{0}(N)$, and charged, 
$\lambda_{-} := E_{0}(N+1) - E_{-}(N+1)$, system.
A sketch representing the calculation procedure is given in Fig. \ref{f3}.

\begin{figure}
        \includegraphics[width=0.95\linewidth]{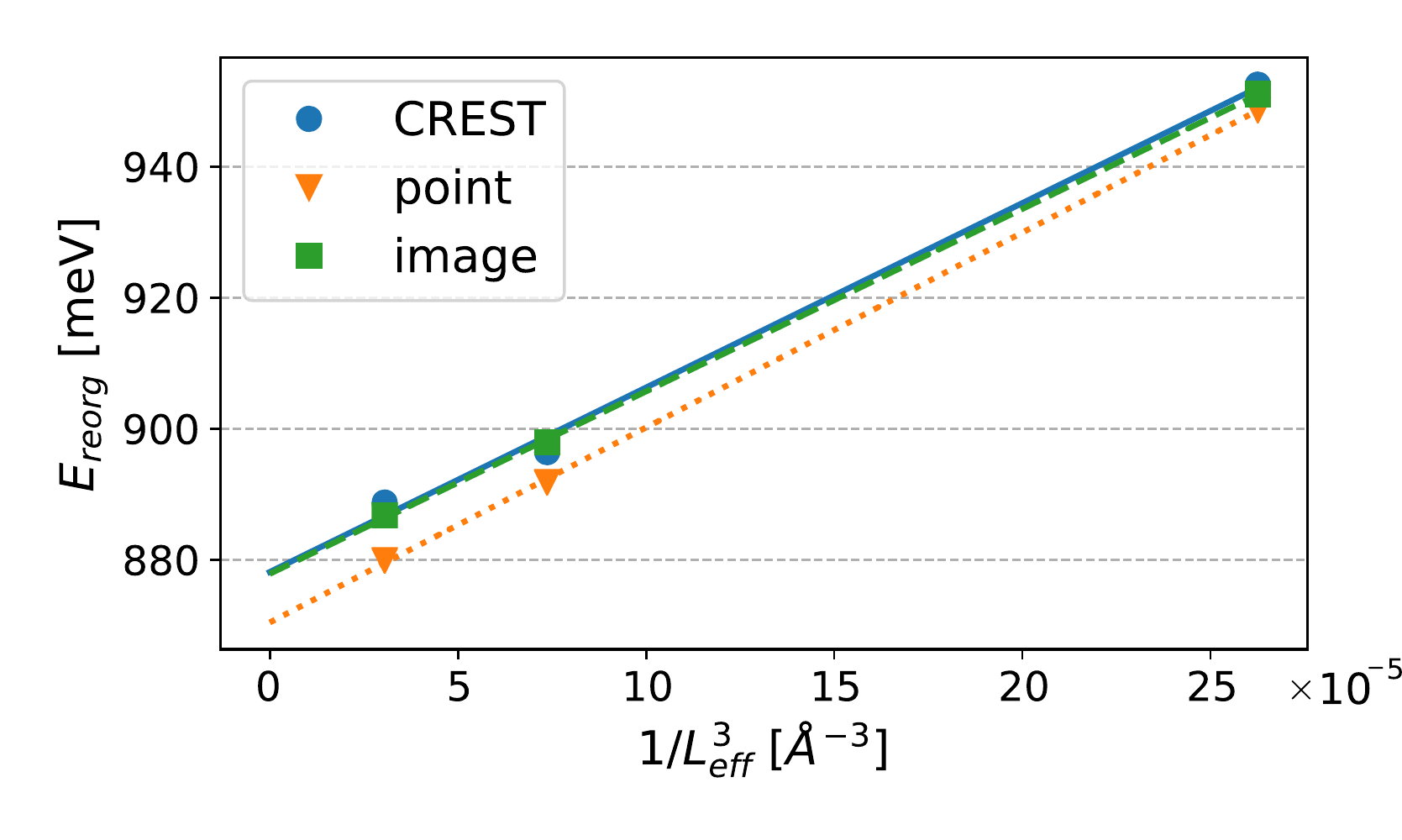}      
    \caption{Reorganization energies as a function of  $1/L_\textnormal{eff}^3$ for CREST (blue), point (orange) and image (green) charge
    compensation methods. The straight lines correspond to the linear fit used to extrapolate the reorganization energy in the infinite-size
    cell limit, $L_\textnormal{eff} \rightarrow +\infty$.
    }\label{f11}
\end{figure}

The polaronic shifts obtained for the different models of 
metal substrate with explicit counter-charges
(CREST, point, shaped) are
summarized in Fig. \ref{f11}. We show $E_\textnormal{reorg}$
as a function of $L_\textnormal{eff}^{-3}$, 
where $L_\textnormal{eff} = \sqrt{L_x, L_y}$
is an effective lateral size (here $L_{x/y}$ 
is the length of the supercell in the $x/y$ direction).
We compare these results
to the polaronic shift computed with the jellium-model {\color{black} in Appendix \ref{app:jellium_vs counter}}.

We note that our calculations operate with periodic replicas of the system and thus do not represent a single molecule but instead a molecular lattice with the supercell size as a lattice constant. 
Due to its long-range nature, the Coulomb interaction introduces interactions - roughly of a dipolar nature - 
between the charged molecule and its periodic replicas. 
Dipolar interactions manifest in 
$E_\textnormal{reorg}$ 
as finite-size corrections that 
converge slowly with increasing the effective 
supercell lateral size.
To explore the single-molecule limit, we extrapolate the reorganization energy using a form $E_\text{reorg}(L_\text{eff}) = a/L_\text{eff}^3 + E_\text{reorg}(\infty)$, 
which is motivated by assuming parallel dipole-dipole interaction.\cite{Jackson}
We extract a polaronic shift, $E_\text{reorg}(\infty)$, for CREST of $878 \pm 9$ meV;
for point charge, $877.8 \pm 0.3$ meV and 
for the shaped image charge distribution, $870.4 \pm 0.4$ meV. %
%
%

%
Similar to the situation of the atomistic structure, we have checked
the importance of including CM corrections into the 
vdW dispersion forces for the reorganization energy.
We find that the scaling with the lateral dipole
correction yields a polaronic shift increased by $\sim 7$\%,
which amounts to $E_\textnormal{reorg}(\infty) = 928.6 \pm 0.1$ meV {\color{black} for the shaped image charge distribution}.
The increase in this value is simply a manifestation of 
the larger ionic displacements after charging due
to larger polarizability of the NaCl surface.

\textcolor{black}{
So far, we have not considered the dependence of the 
reorganization energy on the
number of NaCl layers, $N_l$, which was four in our simulations and 
$\sim 20$ in experiments.\cite{Patera2019} 
From simple electrostatic considerations, and taking into account
that for large $N_l$ the interaction between the charged molecule
and its image charge becomes point-like, we expect
$E_\textnormal{reorg}$ to slowly decay with a dominant
$1/N_l$
functional dependency. We observe, nevertheless, that 
the impact of different models for the electrostatic
interaction between the charged molecule and its image
charge is small, as seen in Fig. \ref{f11}, and therefore,
the impact on the reorganization energy is predicted to 
be quantitatively also small.
Note that as a difference to the lateral dipole correction to the reorganization energy, the dependency 
on the number of layers is not a consequence of periodicity 
of \textcolor{black}{the present single-molecule \textcolor{black}{calculations} 
if the number of NaCl layers would be modified}.
}

\section{Comparison to experiment}

\subsection{Experimental measurement of orbital line shapes}\label{sec:sts}
We turn to the interpretation of the experimental
results reported in Patera \textit{et al.} [\onlinecite{Patera2019}].
In this experiment an individual pentacene molecule adsorbed on thick ($> 20$ monolayers)
NaCl film was repeatedly charged and discharged by tunneling from and to the conductive tip
of an  atomic force microscope.
The Coulomb forces from the charged pentacene act on the oscillating tip
and thereby lead to a measurable additional dissipation of the cantilever.
By measuring this quantity, the spatially resolved probability to induce different charge transitions by tunneling can be mapped out.
With this method, dubbed alternating-charging STM (AC-STM), 
the electronic transitions $0 \rightarrow 1^{-}$
and $1^{-} \rightarrow 0$ have been mapped,
which can be associated to tunneling into \textcolor{black}{the LUMO} and out of the former LUMO.
\textcolor{black}{Here, `former LUMO' refers to the 
singly-occupied orbital of the pentacene anion which derives from the LUMO of neutral pentacene.}
As the tunneling process occurs much faster than ionic relaxations
the former and latter transitions proceed 
in the relaxed geometry of neutral (for $0 \rightarrow 1^{-}$) and charged 
(for $1^{-} \rightarrow 0$) pentacene/NaCl, respectively. 
In the following, these geometries we denote as geo$^0$
and geo$^{-1}$, respectively. 
The slow response to the charge in geo$^{-1}$
then manifests as an increase of the amplitude 
in the central lobe of the signal at the expense of a simultaneous
decrease in the outer lobes (``orbital confinement'') as compared to geo$^{0}$.
These differences were attributed to substrate-ion relaxation after electron transfer (``polaron formation'').

\begin{figure}
        \includegraphics[width=1.0\linewidth]{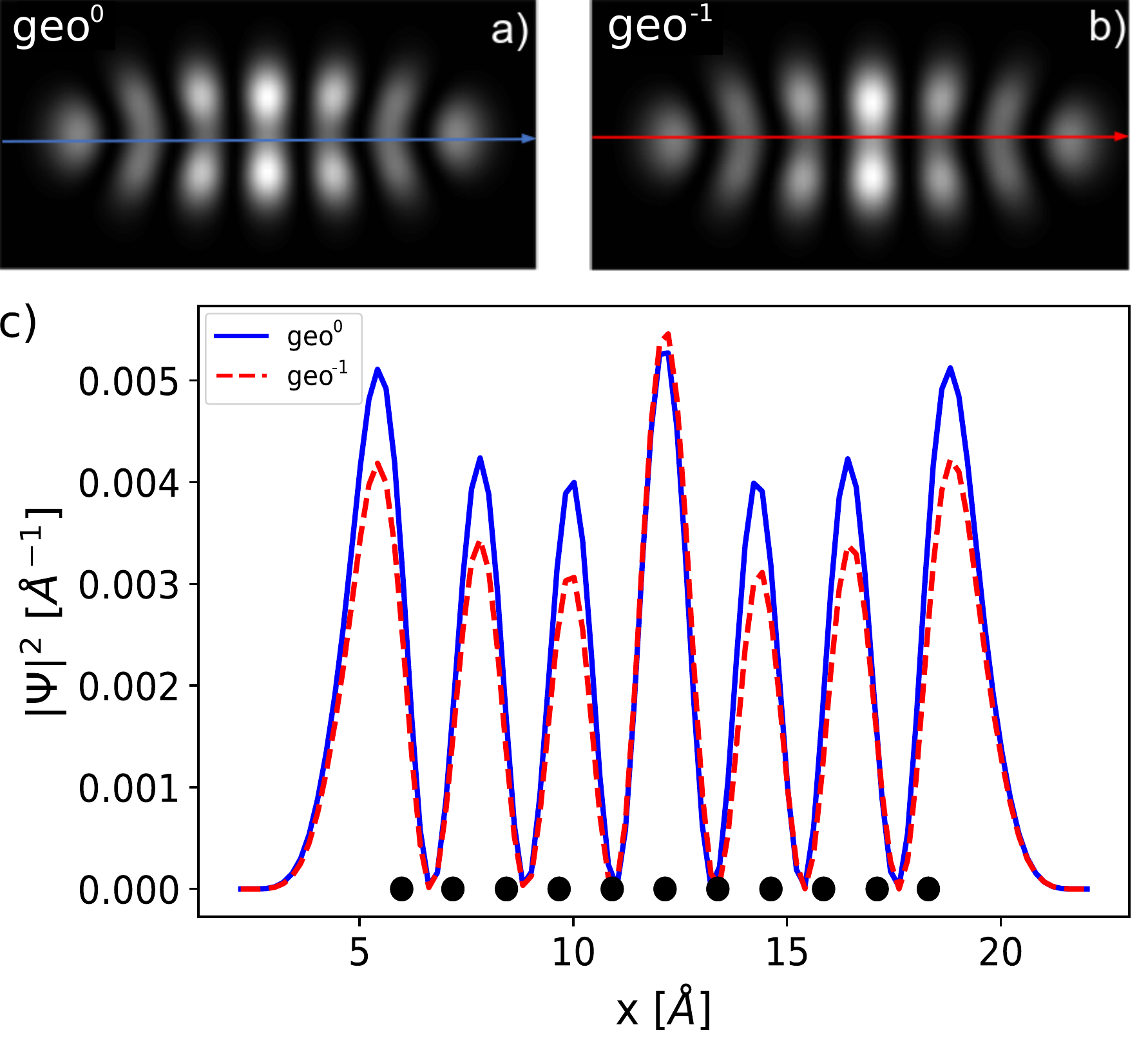}      
    \caption{DFT-calculated probability density for the former LUMO (orbital designation refers to the neutral molecule)
    of charged pentacene/NaCl using 
    (a) neutral geometry and (b) charged geometry. The isosurface
    of the orbitals is cut through a plane located at $\Delta z = 2.4\,\text{\AA}$ above the molecular plane. 
    (c) One-dimensional densities, $|\Psi(x)|^2 \equiv |\Psi(x, y_0, z_0)|^2$, along
    the molecular long symmetry axis indicated in (a) and (b). The position of the carbon atoms
    projected onto this axis is denoted by black spheres.
    }\label{f13} 
\end{figure}

\begin{figure}
        \includegraphics[width=1.0\linewidth]{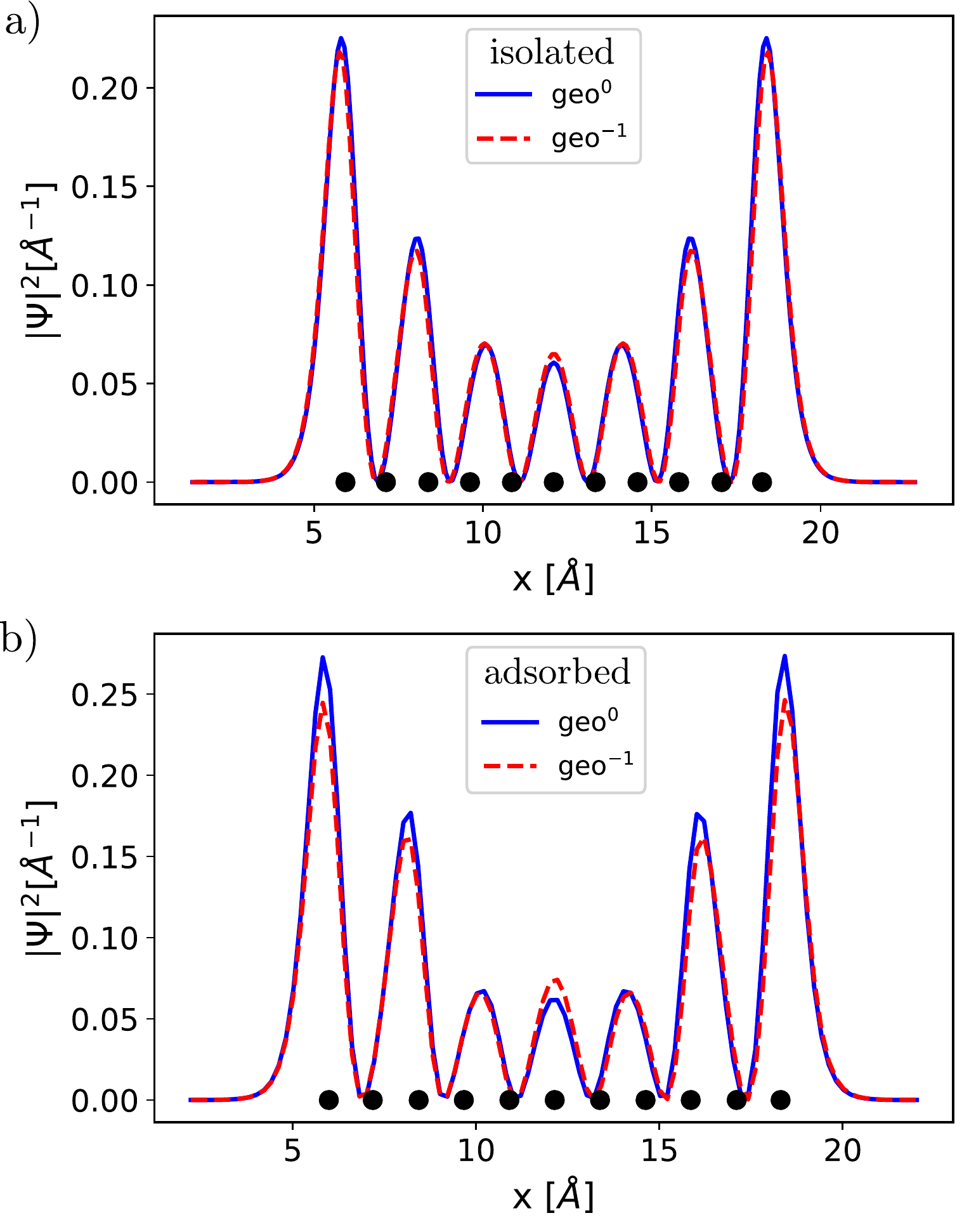}      
    \caption{\textcolor{black}{One-dimensional probability densities, $|\Psi(x)|^2 \equiv |\Psi(x, y_0, z_0)|^2$, of the former LUMO orbital obtained along the lines defined in Fig. \ref{f13}. The lines are contained in a plane located at $\Delta z = 1.0\,\text{\AA}$ above the molecular plane. Here (a) corresponds to the isolated molecule and (b) to the adsorbed molecule; the continuous blue line refers to the neutral geometry while the red dashed line corresponds to the charged geometry. The position of the carbon atoms
    projected onto this axis is denoted by black spheres.  %
    }
    }\label{f15} 
\end{figure}

\subsection{Computational results} 

To simulate the experimental results for polaron formation, 
we first obtain the former LUMO orbital, $\Psi(x,y,z)$, for a
charged pentacene/NaCl system before 
(neutral geometry) and after
charge-induced geometry relaxation
(charged geometry). 
We use the same computational settings as in the previous section 
%
For simplicity, we consider line cuts along the 
main molecular axis (here by convention the $\hat{\mathbf{x}}$ axis)
obtained from the 3D wavefunction,%
$\Psi(x)\equiv \Psi(x, y_0, z_0)$.
We show in Fig. \ref{f13} the probability density of the 
former LUMO (orbital designation refers to the neutral molecule)
for charged pentacene/NaCl with (a) neutral geometry (geo$^0$)
and (b) charged geometry (geo$^{-1}$) along the main molecular axis.
They are both characterized by seven lobes 
and are qualitatively similar to the 
LUMO of uncharged pentacene/NaCl, as expected.
In Fig. \ref{f13} (c) we also show the line cuts, $|\Psi(x)|^2$,
for the geometries in (a) and (b). We note that taking into
account the relaxation of the NaCl film leads to a
decrease of the density in the outer lobes together with a small increase
for the central lobe only. 

\textcolor{black}{
The increase in the amplitude
of the central lobe at the expense of the outer ones (confinement) as a consequence of relaxation has a twofold origin. The first has been discussed 
in Sec. \ref{sec:molecule}: 
the geometry of the molecule changes after charging.  
The second is the substrate reorganization shown in Fig. \ref{f10} that effectively leads to a reshaping of the single-particle potential seen by the molecular charge density. 
}

\textcolor{black}{
To estimate the relative contributions to 
confinement we show in Fig. \ref{f15} the line cuts  $|\Psi(x)|^2$ for (a) the 
isolated and (b) NaCl-adsorbed pentacene molecule. These cuts have been obtained at a distance $\Delta z = 1.0$ \AA\, from
the molecular plane. Although the general trend (orbital confinement) is already present in (a) due to the geometry relaxation of pentacene after charging, we see that this only accounts for approximately 25\% of the change in the relative heights of the maxima in (b). The rest we attribute to  substrate reorganization of the  NaCl lattice, which also produces the large workfunction shifts discussed in Sec. \ref{sec:molecule}.
}

\begin{figure}
        \includegraphics[width=0.95\linewidth]{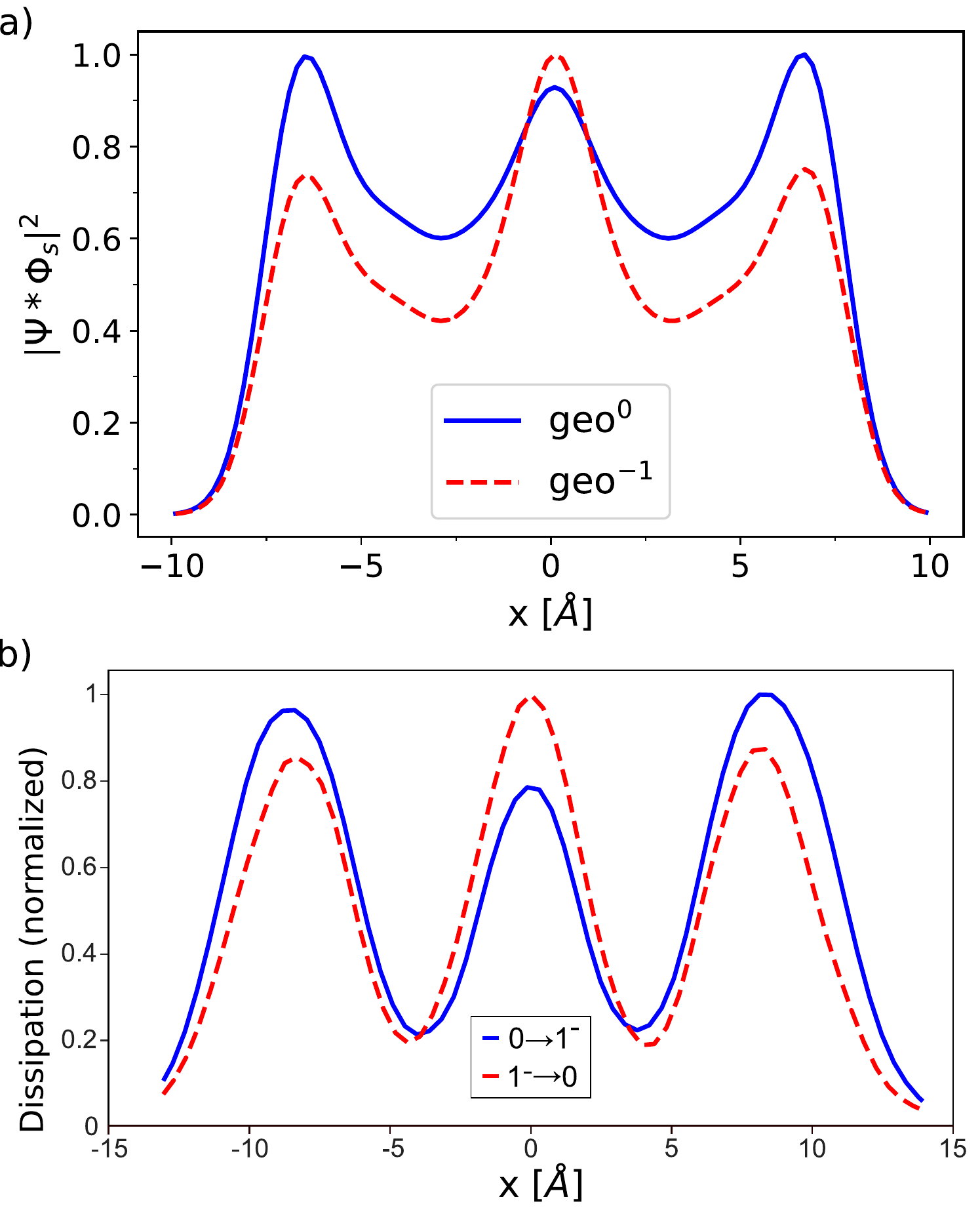}      
    \caption{DFT-simulated (a) and experimental (b) alternating-charging STM signal 
    profile of charged pentacene adsorbed on a NaCl film. The vertical scale is 
    normalized to one.
    Blue solid curves corresponds to
    the neutral geometry, $\textnormal{geo}^0$, and red dashed 
    curve corresponds to the relaxed geometry, $\textnormal{geo}^{-1}$. 
    The experimental signal reflects the spatially resolved 
    possibility to charge (blue) or discharge (red) pentacene/NaCl, measured as an
    additional dissipation signal in alternating-charging STM\cite{Patera2019}.
    }\label{f12} 
\end{figure}

\subsection{Simulation of tip effects}

To simulate effects of the STM tip, we adopt a conventional procedure\cite{Tersoff1985, Chen2007}
and convolve the molecular orbital $\Psi(x)$ with 
a model wavefunction for the STM tip
\begin{equation}\label{eq:convolution}
 [\Psi \ast \Phi_{\textnormal{s}}](x) = \int_{-\infty}^{+\infty}  \Psi(y) \Phi_\textnormal{s}(x - y) \textnormal{d} y.
\end{equation}

We here adopt the approximation 
\begin{equation}\label{eq:s-wave} 
\Phi_{\textnormal{s}}(x) = \exp \left[-\dfrac{x^2}{\sigma^2} \right]
\end{equation}
which should be appropriate for  s-type tip orbitals.
Equation \eqref{eq:convolution} describes the overlap between the tip and molecular orbitals as the tip slides
along the $\hat{\mathbf{x}}$ axis.
In principle, there could be more than one channel (\textit{i.e.}, s-wave, p-wave, \textit{etc}.) with additive contribution to the total STM signal.\cite{Gross2011a} 
We have verified that possible p-wave contributions have a negligible contribution
to the simulated spatial tunneling amplitude. 
This is consistent with the fact that the AC-STM signal acquired with a non-functionalized copper tip comes mostly from $s$-type tip orbitals.

Using Eqs. \eqref{eq:convolution} and \eqref{eq:s-wave} with a Gaussian broadening of $\sigma = 2.5$ \AA, we obtain the simulated AC-STM spatial tunneling amplitude shown in Fig. \ref{f12}. 
As a result of the convolution with the tip orbital, the AC-STM signal 
shows only three lobes.
Before structural relaxation, the central lobe has
smaller amplitude compared to the two outer lobes (geo$^0$).
After relaxation (geo$^{-1}$), the situation is reversed:
The inner lobe increases the amplitude while 
the amplitude for the outer lobes diminishes, in very good agreement with the experiment (cf. Fig.~\ref{f12}).
%

%
\textcolor{black}{
\subsection{Reorganization energy}
The reorganization energy for charge transfer is not just a property of the substance under study 
but strongly depends on the environment. Whereas reorganization energies are routinely measured 
for ensembles, measuring the reorganization energy of 
a single molecule has only recently become experimentally 
accessible~\cite{Fatayer2018}.
By means of atomic force microscopy with single-charge sensitivity 
the \textcolor{black}{hole} reorganization energy of an isolated naphthalocyanine molecule has been determined to 
be \textcolor{black}{$0.8 \pm 0.2$\,eV}. 
These experiments have been conducted for isolated molecules on thick NaCl layers and
hence in the same environment as the one considered here.
While both molecules are planar and feature delocalized frontier orbitals of the $\pi$-system,
naphthalocyanine is slightly larger than pentacene.
We therefore estimate the \textcolor{black}{electron} reorganization energy of pentacene to lie in 
the same range as for naphthalocyanine, namely to be on the order of 0.8-1.0\,eV.
}

\section{Conclusion} 

To recapitulate, we have investigated structural
and energetic properties of NaCl(001) slabs as well as neutral and charged 
individual pentacene molecules adsorbed on NaCl(001) films.
We have analyzed the geometry (mean layer separation
and buckling) both
for thin and thick NaCl slabs and 
found pronounced odd-even effects.
We find a strong dimerization between
Na$^{+}$ and Cl$^{-}$ ions of neighboring 
layers, possessing long range order
that goes deep into the NaCl slab.
We provide a precise calculation of the
surface energy per area
from finite-size scaling
of the energy per layer 
to the bulk limit
and obtain the value \textcolor{black}{$\gamma = 9.6 \pm 0.2\,\, \si{meV}/{\text{\AA}^2}$}.

We have also investigated the 
adsorption properties 
of isolated pentacene/NaCl.
We find that the electrostatic interaction between
the charge density of charged pentacene and the NaCl surface
produces a reversal of the buckling pattern
on the first layers of the NaCl film.
We have checked the consistency of
our results for 
different models of screening at the metal 
substrate that have explicit
counter-charges and investigated
the impact of dispersion forces on
the pentacene/NaCl atomistic structure.
We have calculated the reorganization energy
associated to relaxation of 
the surface upon charging and found it to be
$E_\textnormal{reorg} \simeq 870-930$ meV; depending
on the functional used.
This observable has so far not
been measured for pentacene/NaCl surfaces but the 
value is consistent with the known reorganization
energy of other molecules of similar size adsorbed on NaCl.\cite{Fatayer2018}
Finally, we have studied the spatial changes of the 
STM signal associated to the LUMO.
We have confirmed from \textit{ab-initio} calculations
and simple modeling of the tip-sample interaction
the localization of the charge density upon relaxation \textcolor{black}{of the molecule-insulator interface} as
observed in a recent experiment\cite{Patera2019}.
The experimental results are thus interpreted as being of 
polaronic-like nature.

\begin{acknowledgments}
The authors acknowledge E. Wruss
and O. Hofmann for insightful suggestions. 
Support from the German Research Foundation (DFG) through the Collaborative Research Center,
 Project ID No. 314695032 
SFB 1277 (Projects No. A03 and No. B01) is gratefully acknowledged. 
D. A. E. additionally acknowledges funding from the Alexander von Humboldt Foundation within the framework of the Sofja Kovalevskaja Award, endowed by the German Federal Ministry of Education and Research.
\end{acknowledgments}


\appendix

\section{Bulk NaCl}\label{app:bulk}
In this Appendix, we study the bulk properties of NaCl as 
reference for our investigation of NaCl films.
NaCl is an ionic solid formed by the arrangement of 
two face-centered cubic lattices; 
one formed by negatively charged Cl$^-$ ions and
another by smaller
positively charged Na$^+$ ions. 
The Na/Cl lattices are shifted against each other
by $(1,1,1)\, a_0/2$ where $a_0$ is the lattice constant.
\subparagraph{Results}
Our computed lattice constants are summarized in Table \ref{tS1}.
The lattice constant from our PBE calculations are in 
good agreement to previous theoretical studies\cite{Li2007, Zhang2011, Bucko2010}.
Compared to experimental measurements, the lattice constant 
is overestimated at least $0.06$ \AA\, ($1$\%$a_0$).
Due to the impact of long-range screening on the polarizability of atoms in solids, we also checked the effect
of including vdW dispersion within the TS\cite{Tkatchenko2009}
and TS-CM\cite{Zhang2011} approaches.
The calculated lattice constants 
underestimate the experimental one by at least $0.18$ \AA\, (for PBE-TS)
and $0.065$ \AA\, (for PBE-TS-CM).
This results from overestimation of
the dispersion forces for a bulk solid\cite{Bucko2013}.

\begin{table}
	\centering
			\setlength{\tabcolsep}{10pt} 

		\begin{tabular}[b]{l|c c}
		\toprule
		& $a_0$ [\AA]	& $E_\textnormal{G}$ [eV] 	\\
		\midrule
		PBE						& 5.700 		& 5.03 			\\
		PBE-TS 					& 5.389 		& 5.66 			\\
		PBE-TS-CM 				& 5.505 		& 5.39 			\\
		PBE [\onlinecite{Li2007}]		& 5.70			& 5.0 			\\
		Exp. 					& 5.57~\cite{Wang1996} - 5.64~\cite{CRCHandbook} 	& 8.5 (optical)~\cite{CRCHandbook} \\
		\bottomrule
	\end{tabular}
	\caption{Lattice constants, $a_0$, and band-gap energies, $E_G$, of bulk NaCl 
	obtained using PBE, and van der Waals corrected
	PBE-TS and PBE-TS-CM functionals. The accuracy in the determination of
	the lattice constant is $0.1$ pm and \textcolor{black}{the band gap is $0.01$ eV.}
	For comparison, DFT results from Ref. \onlinecite{Li2007} and experimental values from Refs. \onlinecite{Wang1996, CRCHandbook} are shown.}
	\label{tS1}
\end{table}

\begin{figure}
        \includegraphics[width=0.95\linewidth]{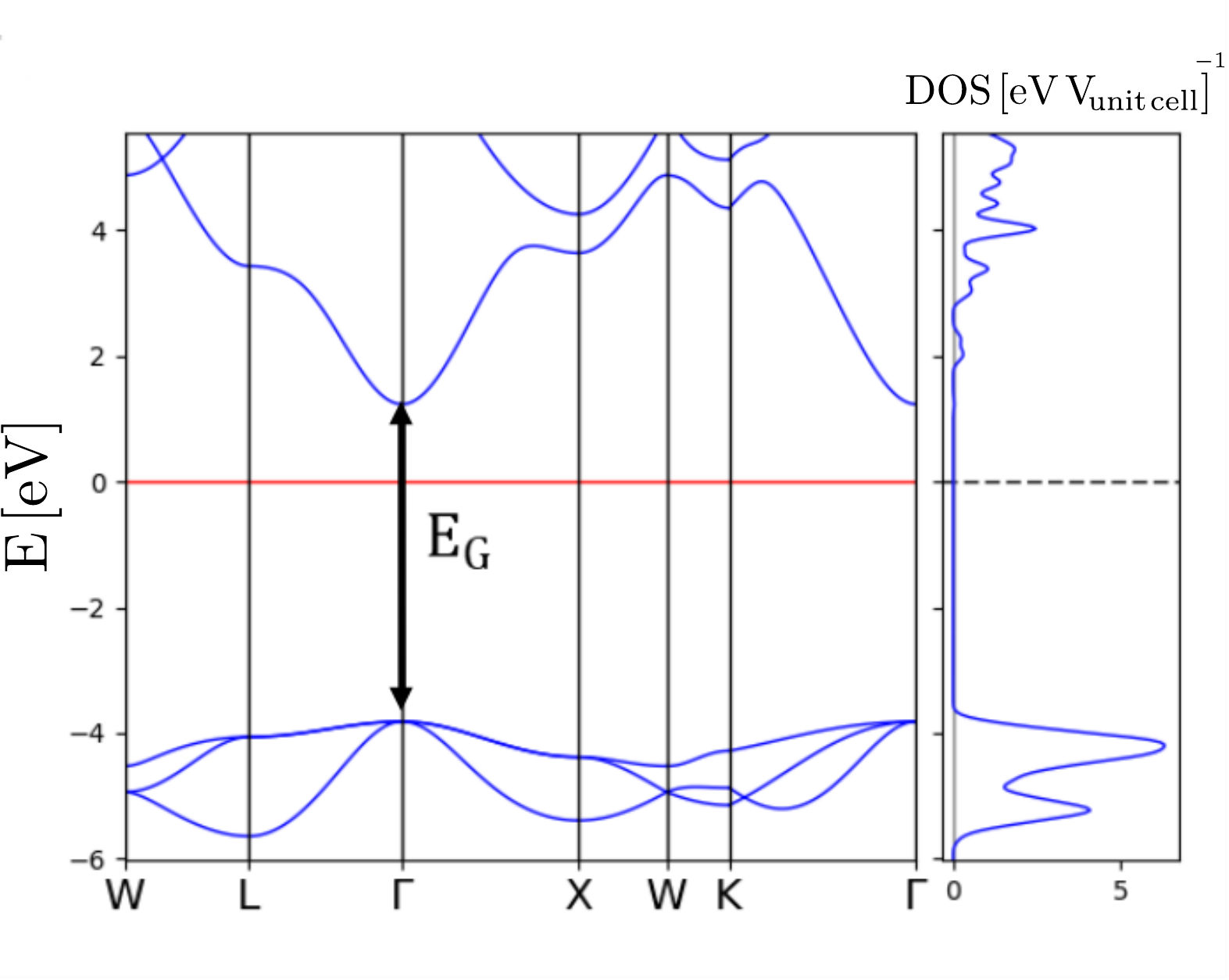}      
    \caption{Band structure and DoS of bulk NaCl obtained using the 
    PBE functional. The band gap at the $\Gamma$ point is denoted by $E_\textnormal{G}$.
    The red horizontal line corresponds to the reference energy within the band gap.
    }\label{fS5} 
\end{figure}

We show in Fig. \ref{fS5} the band structure of NaCl obtained using the PBE functional (similar band structures are obtained for PBE-TS
and PBE-TS-CM here not shown).
The band gap energy for different functionals obtained at the $\Gamma$
point is summarized in Table \ref{tS1}.
This band gap strongly underestimates 
the experimental optical gap $\sim 8.5$ eV \cite{Lipari1971} due to well-known artifacts of local and semilocal functionals.\cite{Perdew1985}
For further details on the NaCl band structure, we 
refer the reader to Ref. \onlinecite{Li2007}.

\subparagraph*{Computational details.}
For bulk NaCl structural optimizations, we have used the FHI-aims ``tight'' 
settings\cite{Blum2009} (roughly equivalent to ``double-zeta plus polarization''
quality) and relaxed the unit cell until the residual
value for every component of the force acting in each ion was 
smaller than $0.005$ eV/\AA. We considered a dense $\mathbf{k}$-point
mesh of at least $8\times 8 \times 8$ sampling 
in each direction of the Brillouin zone.

\section{Convergence checks}\label{app:checks}
\begin{figure}
        \includegraphics[width=0.95\linewidth]{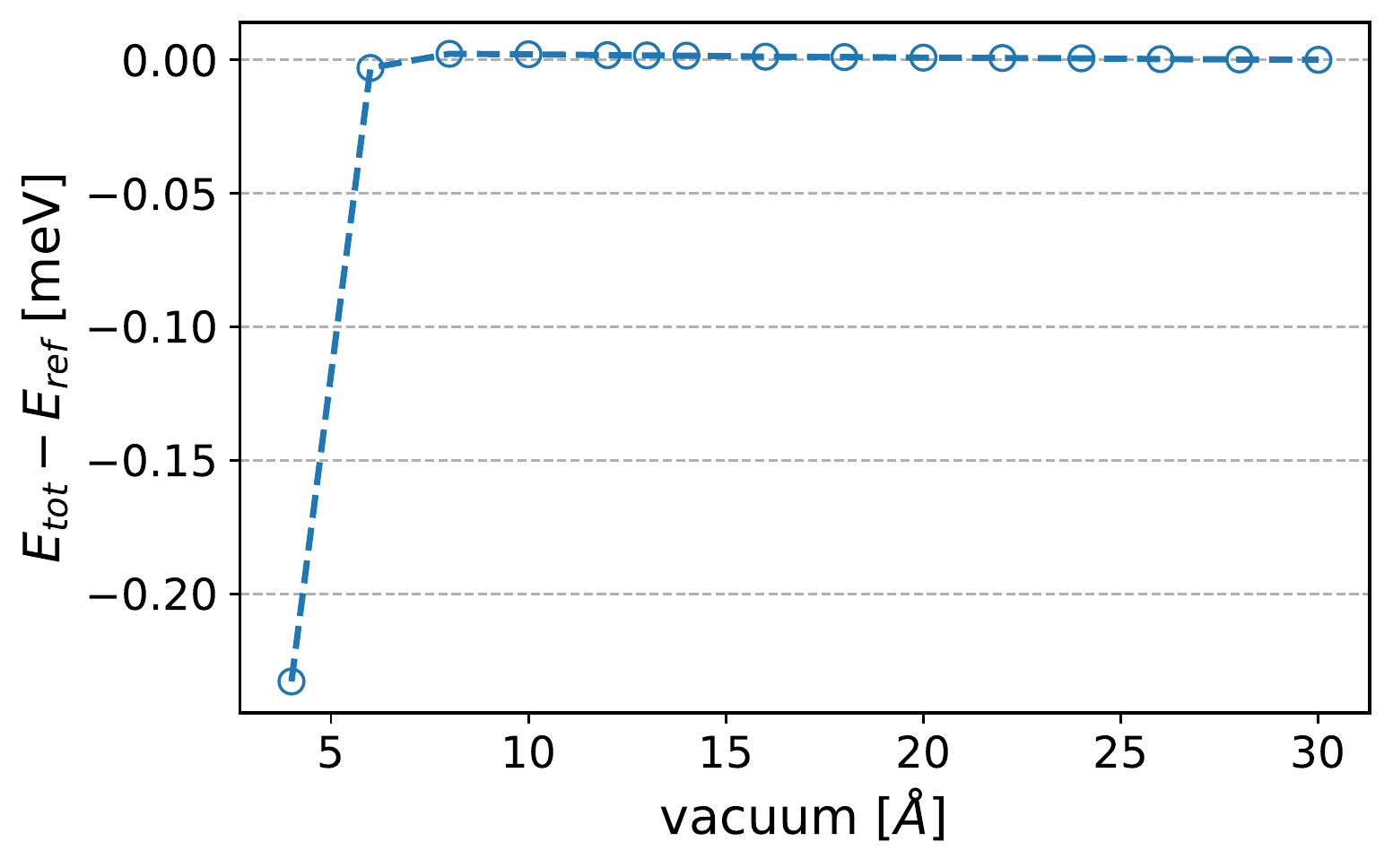}      
    \caption{Variation of the total energy of a 12-layer NaCl(001) slab
    with respect to vacuum layer thickness.
    The energies are normalized to the reference energy, $E_\textnormal{ref} = -204072.096449004$ eV,
    corresponding to the total energy for $30$ \AA \, vacuum.
    }\label{fSC1} 
\end{figure}

\begin{figure}
        \includegraphics[width=0.95\linewidth]{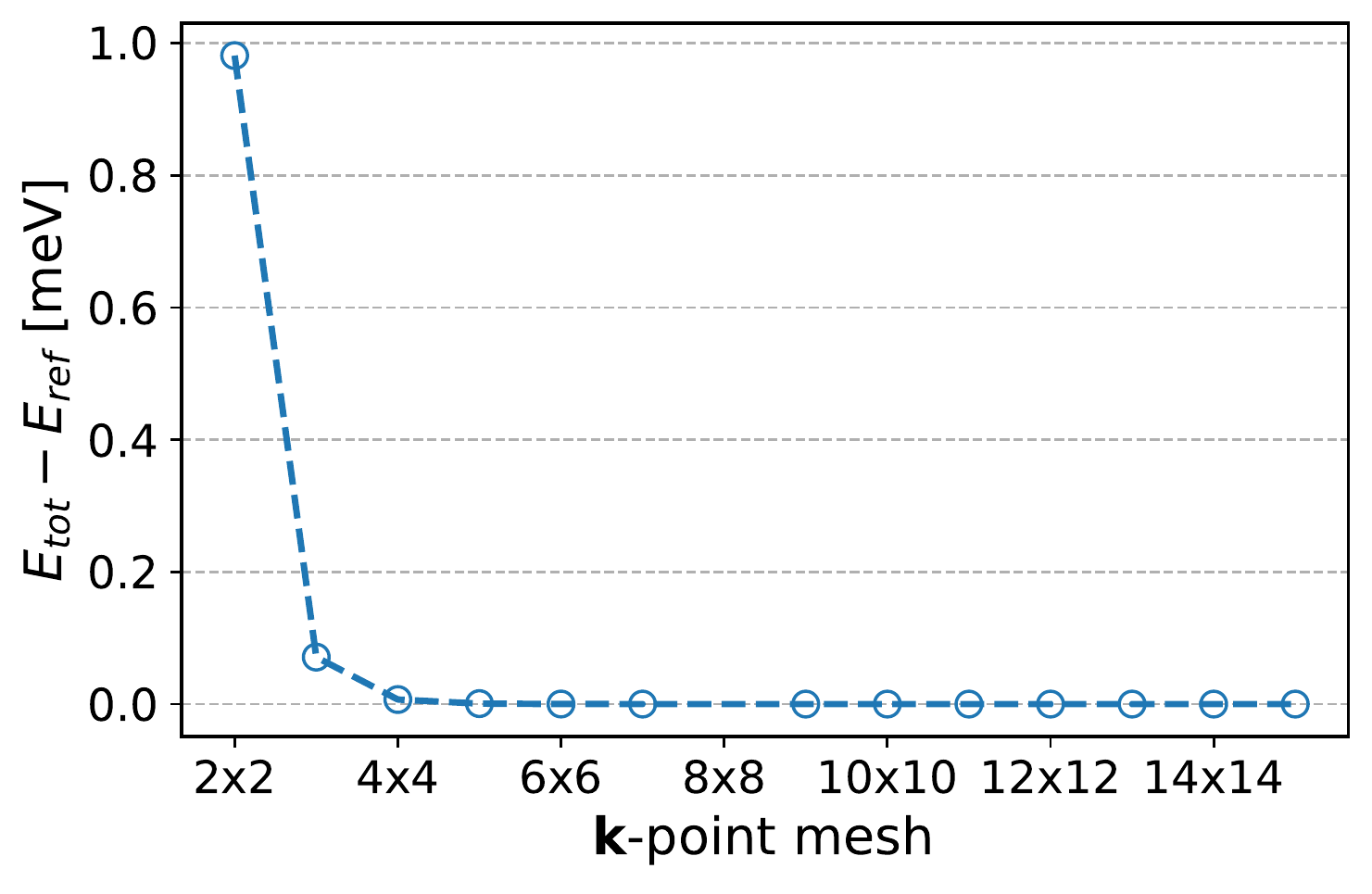}      
    \caption{Variation of the total energy of a 12-layer NaCl(001) slab
    with respect to $k$-point discretization.
    The energies are normalized to the reference energy, $E_\textnormal{ref} = -203227.951979871$ eV,
    corresponding to the total energy for $k$-mesh of $15\times 15$.
    }\label{fSC2} 
\end{figure}

We report some of the checks used to determine the convergence of
our computational setup for the slab calculations. Specifically, we show the dependence of the 
total energy on the width of the vacuum layer in Fig. \ref{fSC1} and 
the and $k$-grid discretization in Fig. \ref{fSC2}
for a slab of
width $L = 12$. 
Similar convergence tests were carried over for slabs of different widths.
%

%

\section{Polaronic shift: jellium vs explicit charge-compensation}\label{app:jellium_vs counter}

\begin{figure}
        \includegraphics[width=1.0\linewidth]{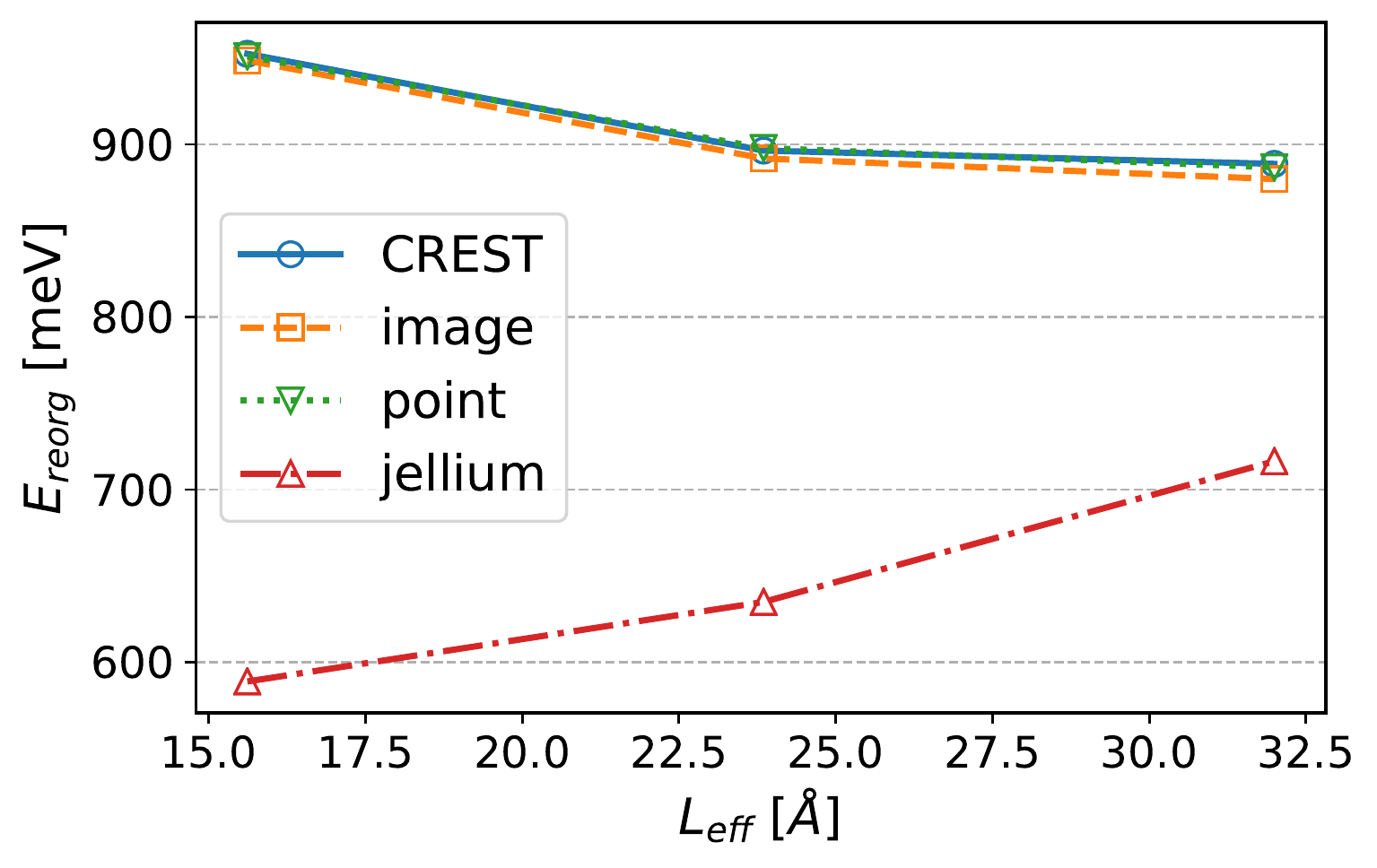}      
    \caption{Reorganization energies as a function of the
    effective lateral size of the supercell, $L_\textnormal{eff} = \sqrt{L_x L_y}$. The four models for the screening due to the metal surface
    from Fig. 2 in the main paper are shown.
    }\label{fS2} 
\end{figure}

In Fig. \ref{fS2}, we compare the reorganization energy computed using
the jellium model and different shaped-counter charge
distributions in the supercell (see main paper). 
As reported in the main text, the explicit counter-charge
models for the metallic surface show a convergence with the lateral
supercell size. However, no convergence is obtained
for the jellium model for which we also report
reorganization energies $200-300$ meV smaller.
The lack of convergence with $L_\textnormal{eff}$ is a consequence of the failures of the jellium model to describe properly electrostatic long range Coulomb interactions in charged supercells with vacuum.

%

\end{document}